\documentclass[amsmath,amssymb,aps,prb,reprint,superscriptaddress,nofootinbib]{revtex4-2}

\usepackage{comment}
\usepackage{enumitem}
\setlist[enumerate]{itemsep=0mm}
\usepackage{color}
\usepackage{graphicx}
\usepackage{braket}
\usepackage{bbm}
\usepackage[outline]{contour}
\usepackage{dsfont}
\usepackage[colorlinks=true,linkcolor=blue,citecolor=blue,urlcolor=blue]{hyperref}

\newcommand{\ee}{{\rm e}}
\newcommand{\ii}{{\rm i}}
\newcommand{\dd}{{\rm d}}

\begin{document}

\title{Slow dynamics from a nested hierarchy of frozen states}

\author{Vanja Mari\'c}%
\affiliation{%
Department of Physics, Faculty of Mathematics and Physics, University of Ljubljana, Jadranska 19, Ljubljana SI-1000, Slovenia
}
\author{Luka Paljk}%
\affiliation{%
Department of Physics, Faculty of Mathematics and Physics, University of Ljubljana, Jadranska 19, Ljubljana SI-1000, Slovenia
}
\author{Lenart Zadnik}
\email{lenart.zadnik@fmf.uni-lj.si}
\affiliation{%
Department of Physics, Faculty of Mathematics and Physics, University of Ljubljana, Jadranska 19, Ljubljana SI-1000, Slovenia
}

\begin{abstract}
We identify the mechanism of slow heterogeneous relaxation in quantum kinetically constrained models (KCMs) in which the potential energy strength is controlled by a coupling parameter. The regime of slow relaxation includes the large-coupling limit. By expanding around that limit, we reveal a \emph{nested hierarchy} of states that remain frozen on time scales determined by powers of the coupling. The classification of such states, together with the evolution of their Krylov complexity, reveals that these time scales are related to the distance between the sites where facilitated dynamics is allowed by the kinetic constraint. While correlations within frozen states relax slowly and exhibit metastable plateaus that persist on time scales set by powers of the coupling parameter, the correlations in the rest of the states decay rapidly. We compute the plateau heights of correlations across all frozen states up to second-order corrections in the inverse coupling. Our results explain slow relaxation in quantum KCMs and elucidate dynamical heterogeneity by relating the relaxation times to the spatial separations between the active regions. 
\end{abstract}

\maketitle

\section{Introduction} 

The rapid progress of quantum technologies provides strong motivation for theoretical investigations into the mechanisms of slow relaxation in quantum many-body systems. A notable example is the growing interest in quantum kinetically constrained models (KCMs), whose classical analogs have historically played an important role in understanding dynamical slowdown and heterogeneity in structural glasses~\cite{palmer-etal-1984,fredrickson-andersen-1984,ritort-sollich-2003,garrahan-2018}. One of the appeals of  quantum KCMs lies in the fact that they are experimentally accessible within Rydberg-atom platforms~\cite{urban-etal-2009,lesanovsky-2011,lesanovsky-garrahan-2013,bernien-etal-2017,bluvstein-etal-2021,kim-etal-2024,yang-etal-2025,corcoran-etal-2025}. From a theorist's standpoint, these systems are remarkable not only because they exhibit slow relaxation, but also because they support a rich variety of other exotic nonequilibrium phenomena. These range from Hilbert space fragmentation~\cite{yang-etal-2020,langlett-xu-2021,zadnik-fagotti-2021,pozsgay-etal-2021,tamura-katsura-2022} and quantum many-body scars~\cite{kerschbaumer-etal-2025,moudgalya-bernevig-ragnault-2022,serbyn-abanin-papic-2021,surace-2021} to anomalous transport properties~\cite{singh-etal-2021,yang-2022,mccarthy-etal-2025,bhakuni-etal-2025}.

In this work we focus on the emergence of slow relaxation in quantum KCMs with a coupling parameter that tunes the strength of the potential energy. Specifically, we are interested in KCMs with two dynamical regimes, depending on the coupling strength: one exhibiting fast correlation decay and another where relaxation is markedly slowed, exhibiting metastable behavior~\cite{horssen-levi-garrahan-2015,lan-horssen-powell-garrahan-2018,feldmeier-pollmann-knap-2019,roy-lazarides-2020,pancotti-giudice-cirac-garrahan-banuls-2020,zadnik-garrahan-2023,causer-banuls-garrahan-2024,menzler-banuls-meisner-2025}. The latter regime is associated with the emergence of a hierarchy of (prethermal) plateaus in the correlation functions. Such a dynamical slowdown has been addressed through the lens of emergent quantum many-body scars and enhanced kinetic constraints in the strong-coupling limit~\cite{pancotti-giudice-cirac-garrahan-banuls-2020,zadnik-garrahan-2023,causer-banuls-garrahan-2024}, as well as emergent (quasi)conserved quantities~\cite{abanin-deroeck-ho-huveneers-2017}. Here, we go beyond the strong-coupling limit by determining the microscopic mechanism that pinpoints the origin of \emph{all} plateaus, thus explaining the entire hierarchy of relaxation times. 

Specifically, we develop a framework within which both the time scales and the heights of the plateaus can be determined. Based on the full large-coupling expansion of the Hamiltonian, our framework defines a nested hierarchy of computational-basis states that are preserved by a sequence of effective Hamiltonians governing the dynamics on different time scales. The latter scale exponentially with the distances between the active regions of space within which the dynamics is allowed under the kinetic constraint. 

We will focus on the so-called XPX model~\cite{zadnik-garrahan-2023,fagotti-maric-zadnik-2024,eck-fendley2024,cao-miao-yamazaki-2025}, but the results are applicable to other quantum KCMs. In support of this claim, we present additional results for the quantum version of the Fredrickson-Andersen model~\cite{fredrickson-andersen-1984,ritort-sollich-2003,hickey-genway-garrahan-2016} in Appendix~\ref{app:FA-model}.

\section{Slow dynamics in the XPX model} 

We consider the one-dimensional XPX model
\begin{align}
    H_{\rm XPX}=\sum_{j=1}^L\sigma^x_{j-1}(\mathds{1}-\sigma^z_j)\sigma^x_{j+1}+\Delta \sigma^z_j
    \label{eq:XPX-model}
\end{align}
with periodic boundary conditions. Here, $\sigma_j^\alpha$, for $\alpha\in\{x,y,z\}$, are Pauli matrices in site $j$. The coupling parameter $\Delta$ controls the strength of the potential energy $V=\sum_{j=1}^L\sigma_j^z$. The kinetic term is constrained: spins $\downarrow$ facilitate simultaneous spin flips on the neighboring sites.

\begin{figure}[t!]
    \centering
    \includegraphics[width=0.485\textwidth]{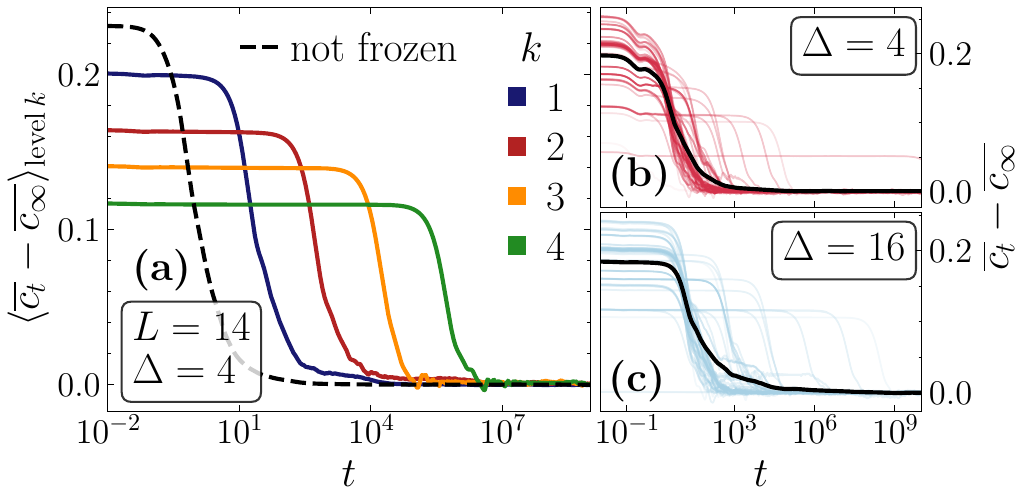}
    \caption{
        {\bf Slow relaxation in the XPX model.} Panel (a) shows a hierarchy of correlations $\overline{c_t}(\boldsymbol{s})=t^{-1} \int_0^t\dd \tau c_\tau(\boldsymbol{s})$, Eq.~\eqref{eq:correlation}, averaged over the computational-basis states $\ket{\boldsymbol{s}}$, frozen on time scales $t\sim\Delta^k$ (the so-called ``level-$k$ states''---see Fig.~\ref{fig:hierarchy}). The correlation function averaged over the rest of the states instead decays rapidly (dashed line). Panels (b) and (c) show correlations in all frozen states, as well as their average (black line). The initial plateau is more pronounced at larger $\Delta$. In all plots the system size is $L=14$.}
    \label{fig:slow-relaxation-xpx}
\end{figure}

Slow dynamics arises for $|\Delta|>1$~\cite{zadnik-garrahan-2023} and it can be observed in the behavior of time-averaged correlation functions $\overline{c_t}(\boldsymbol{s})=t^{-1}\int_0^t \dd\tau c_\tau(\boldsymbol{s})$, where
\begin{align}
    c_t(\boldsymbol{s}) := \frac{1}{L}\sum_{j=1}^L\bra{\boldsymbol{s}}n_j(t)n_j(0)\ket{\boldsymbol{s}}.
    \label{eq:correlation}
\end{align}
Time averaging is employed to smooth out quantum coherences on short time scales. Here and in the following, $\ket{\boldsymbol{s}}\equiv\ket{s_1\ldots s_L}$ is a computational-basis state, with $s_j\in\{\uparrow,\downarrow\}$ labeling the eigenstates of $\sigma_j^z$, and $n_j(t)=\ee^{\ii t H_{\rm XPX}}n_j \ee^{-\ii t H_{\rm XPX}}$ is the Heisenberg time evolution of the site occupancy $n_j=(\mathds{1}+\sigma^z_j)/2$. In Figs.~\hyperref[fig:slow-relaxation-xpx]{1(b)} and~\hyperref[fig:slow-relaxation-xpx]{1(c)} we show a selection of slowly relaxing correlation functions in the $|\Delta|>1$ regime. For larger $\Delta$, the first among the plateaus becomes more pronounced, persisting up to times $t\sim \Delta$~\cite{zadnik-garrahan-2023,causer-banuls-garrahan-2024}. In that (large-coupling) regime, the dynamics is governed by an effective Hamiltonian that is more constrained than $H_{\rm XPX}$, its Hilbert space splitting into exponentially many dynamically disconnected sectors~\cite{fagotti-2014,zadnik-fagotti-2021,zadnik-bidzhiev-fagotti-2021,pozsgay-etal-2021,yang-etal-2020}. As a consequence, relaxation is suppressed up to times $t\sim \Delta$, around which corrections to the large-coupling limit become relevant, causing the plateau's decay~\cite{causer-banuls-garrahan-2024,fagotti-2014,zadnik-fagotti-2021}. We will show that this decay occurs in stages and identify the precise mechanism behind such a hierarchical relaxation.

We will start with the key ingredient---the full large-coupling expansion of $H_{\rm XPX}$. Successive truncations of this expansion yield kinetically constrained effective Hamiltonians that govern the dynamics on progressively longer time scales. Due to kinetic constraints, each one of them has a set of computational-basis eigenstates, which we will classify and count. By tracking the evolution of their Krylov complexity~\cite{balasubramanian-etal-2022,nandy-etal-2025} under the full Hamiltonian $H_{\rm XPX}$, we will show that such states are effectively frozen up to times that scale as powers of $\Delta$. The plateaus appear only in the correlation functions in frozen states---they are shown in Fig.~\ref{fig:slow-relaxation-xpx}. We will calculate their heights up to $O(\Delta^{-2})$ corrections.

\section{Large-coupling expansion} 

Our starting point is the recursive scheme of Ref.~\cite{macdonald-1988}, which allows us to compute the large-coupling expansion up to any order in $1/\Delta$ (see Appendix~\ref{app:large-coupling}). It results in an anti-Hermitian $S = \sum_{n\ge 1}\Delta^{-n} S_n$ and a Hermitian $H_{\rm F}=\sum_{n\ge 0}\Delta^{-n} H_{\rm F\!,n}$, such that
\begin{align}
    H_{\rm XPX}\!=\!\ee^{-S}H\ee^{S},\quad H\!=\!H_{\rm F}\!+\!\Delta V,\quad [H_{\rm F},V]\!=\!0,
    \label{eq:large-coupling-expansion}
\end{align}
where $V=\sum_{j=1}^L\sigma_j^z$ is the potential energy.
We will refer to $H$ as the {\em effective Hamiltonian}, and to $H_{\rm F}$ as the {\em folded XPX model}~\cite{zadnik-fagotti-2021,pozsgay-etal-2021}. The name comes from the fact that the spectra of $H_{\rm XPX}$ and $H_{\rm F}$ are equivalent up to shifts that are integer multiples of $\Delta$ (the spectrum of $H_{\rm XPX}$ ``folds'' into the one of $H_{\rm F}$). We also define truncations of the generator and of the effective Hamiltonian,
\begin{align}
    S^{(k)}:= \sum_{n=1}^k\Delta^{-n}S_n, \quad H^{(k)}:= \Delta V+\sum_{n=0}^{k-1}\Delta^{-n}H_{{\rm F}\!,n}. 
    \label{eq:truncations}
\end{align}

The first few orders in the expansion of $H_{\rm F}$ read
\begin{align}
    H_{\rm F\!,0}\!=&\sum_{j=1}^L\frac{\mathds{1}\!-\!\sigma_j^z}{2}(\sigma_{j-1}^x\sigma_{j+1}^x\!+\!\sigma_{j-1}^y\sigma_{j+1}^y),
    \label{eq:H-0}\\
    H_{\rm F\!,1}\!=&
    \sum_{j=1}^L \frac{\mathds{1}\!-\!\sigma_{j-1}^z}{2}\sigma_j^z\frac{\mathds{1}\!-\!\sigma_{j+1}^z}{2}\left(\sigma_{j-2}^+\sigma_{j+2}^- \!+\!{\rm H.c.}\right)\notag\\
    &\!+\!\left(\sigma_{j-1}^+\sigma_{j}^-\sigma_{j+1}^+\sigma_{j+2}^-\!+\!{\rm H.c.}\right)+\sigma_j^z \frac{\mathds{1}\!-\!\sigma_{j+1}^z}{2},
    \label{eq:H-1}
\end{align}
and
\begin{widetext}
\begin{align}
    H_{\rm F\!,2}\!=
    &\sum_{j=1}^L \frac{3}{4}\left(\frac{\sigma^z_{j-2}\!+\!\sigma^z_{j+2}}{2}\!-\!\mathds{1}\right)\!\left(\frac{\mathds{1}\!-\!\sigma_{j}^z}{2}\right)\!\left(\sigma_{j-1}^+\sigma_{j+1}^-\!+\!{\rm H.c.}\right)\!+\!\frac{1}{2}\left(\frac{\mathds{1}\!+\!\sigma_{j}^z}{2}\right)\!\left(\sigma^+_{j-2}\sigma^-_{j-1}\sigma^+_{j+1}\sigma^-_{j+2}\!+\!{\rm H.c.}\right)\notag\\
    &\!-\!\frac{1}{2}\left(\frac{\mathds{1}\!-\!\sigma_{j}^z}{2}\right)\!\left(\sigma^+_{j-2}\sigma^-_{j-1}\sigma^-_{j+\!1}\sigma^+_{j+2}\!+\!{\rm H.c.}\right)\!+\!\frac{3}{4}\left(\frac{\mathds{1}\!-\!\sigma_{j-1}^z}{2}\sigma^z_j\right)\!\left(\sigma_{j-2}^+\sigma^-_{j+1}\sigma^+_{j+2}\sigma^-_{j+3}\!+\!{\rm H.c.}\right)\notag\\
    &\!+\frac{3}{4}\left(\sigma^z_{j+1}\frac{\mathds{1}\!-\!\sigma_{j+2}^z}{2}\right)\!\left(\sigma_{j-2}^+\sigma^-_{j-1}\sigma^+_j\sigma^-_{j+3}\!+\!{\rm H.c.}\right)\!+\!\frac{1}{4}\left(\sigma^z_{j+1}\frac{\mathds{1}\!-\!\sigma_{j+2}^z}{2}\right)\!\left(\sigma^+_{j-2}\sigma^+_{j-1}\sigma^-_j\sigma^-_{j+3}\!+\!{\rm H.c.}\right)\notag\\
    &\!+\!\frac{1}{4}\left(\frac{\mathds{1}\!-\!\sigma_{j-1}^z}{2}\sigma^z_j\right)\!\left(\sigma^+_{j-2}\sigma^+_{j+1}\sigma^-_{j+2}\sigma^-_{j+3}\!+\!{\rm H.c.}\right)\!+\!\left(\frac{\mathds{1}\!-\!\sigma_{j-2}^z}{2}\sigma^z_{j-1}\frac{\mathds{1}\!-\!\sigma^z_{j}}{2}\sigma^z_{j+1}\frac{\mathds{1}\!-\!\sigma^z_{j+2}}{2}\right)\!\left(\sigma^+_{j-3}\sigma^-_{j+3}\!+\!{\rm H.c.}\right),
    \label{eq:H-2}
\end{align}
\end{widetext}
where $\sigma_j^\pm=(\sigma^x_j\pm\ii\sigma^y_{j})/2$. Additionally, we report the leading order of $S$:
\begin{align}
    S_1\!=\frac{1}{2}\sum_{j=1}^L\frac{\mathds{1}\!-\!\sigma_{j}^z}{2}\left(\sigma_{j-1}^+\sigma_{j+1}^+\!-\!\sigma_{j-1}^-\sigma_{j+1}^-\right).
    \label{eq:S-1}
\end{align}
We note that $H_{{\rm F}\!,0}$ is an integrable model, solved exactly in Refs.~\cite{zadnik-fagotti-2021,zadnik-bidzhiev-fagotti-2021,zadnik-bidzhiev-fagotti-2021,pozsgay-etal-2021}, while higher orders of the folded model, $H_{{\rm F}\!,n}$, for $n\ge 1$, break integrability.

\section{Nested hierarchy of frozen states} 

\begin{figure}
    \centering
    \includegraphics[width=0.9\linewidth]{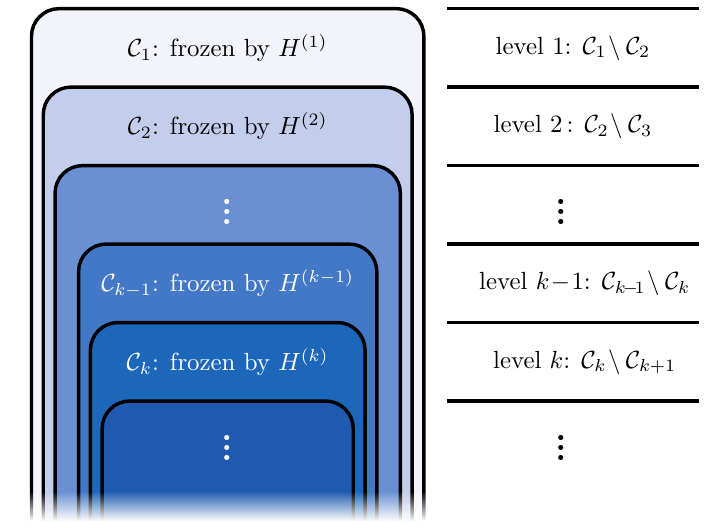}
    \caption{{\bf Nested hierarchy of states.} $\mathcal{C}_k$ contains the states frozen by the truncation $H^{(k)}$ of the effective Hamiltonian $H$. Level-$k$ states are those in $\mathcal{C}_k$ that are only frozen by $H^{(\ell)}$ for $\ell\le k$.}
    \label{fig:hierarchy}
\end{figure}

Due to kinetic constraints in $H_{{\rm F}\!,n}$, each truncation $H^{(k)}$ of the effective Hamiltonian possesses a set of computational-basis eigenstates referred to as \emph{frozen states}\footnote{Since we demand the frozen states to be product states in the computational basis, they are also the eigenstates of the local density $h^{(k)}_j$, where $H^{(k)}=\sum_{j=1}^L h^{(k)}_j$.}:
\begin{align}
    \mathcal{C}_k\!:=\!\big\{\ket{\boldsymbol{s}}\equiv\ket{s_1\ldots s_L} \,\big|\, H^{(k)}\!\ket{\boldsymbol{s}}\!=\!\varepsilon(\boldsymbol{s})\!\ket{\boldsymbol{s}}\big\}.
    \label{eq:frozen-states}
\end{align}
From the way they are defined, such states do not depend on $\Delta$. Since $H^{(k+1)}=H^{(k)}+\Delta^{-k} H_{{\rm F}\!,k}$, according to Eq.~\eqref{eq:truncations}, $H^{(k+1)}$ contains both the constrained hopping terms of $H_{{\rm F}\!,k}$, as well as those already present within the lower-order truncation $H^{(k)}$. Higher-order truncations may allow for hopping processes within spin configurations that were frozen by the lower-order truncations. As a result  $\mathcal{C}_{k+1}\subseteq\mathcal{C}_k$: only those frozen states of $H^{(k)}$ that remain eigenstates in the presence of additional constrained hopping processes described by $H_{{\rm F}\!,k}$ are also frozen states of $H^{(k+1)}$. Frozen states can therefore be classified within a {\em nested hierarchy} of sets
\begin{align}
    \mathcal{C}_1\supseteq\mathcal{C}_2\supseteq\ldots \supseteq \mathcal{C}_{k-1}\supseteq \mathcal{C}_k\supseteq\ldots\, ,
    \label{eq:nested-hierarchy}
\end{align}
depicted in Fig.~\ref{fig:hierarchy}.

For illustration, consider first $H^{(1)}=\Delta V+H_{{\rm F}\!,0}$, where the constraint comes from the leading-order folded model in Eq.~\eqref{eq:H-0}. The only nontrivial hopping process allowed by the latter is $\ket{\cdots\uparrow\downarrow\downarrow\cdots}\leftrightarrow \ket{\cdots\downarrow\downarrow\uparrow\cdots}$: it requires the presence of a pair of neighboring $\downarrow$-spins. $\mathcal{C}_1$ therefore contains spin configurations in which subsequences $\downarrow\downarrow$ are forbidden. Additionally, according to the definition in Eq.~\eqref{eq:frozen-states}, the ``vacuum'' state $\ket{\downarrow\downarrow\cdots\downarrow}$, with all spins $\downarrow$, is trivially contained within $\mathcal{C}_1$.

To identify $\mathcal{C}_2$, i.e., the set of frozen states of $H^{(2)}=H^{(1)}+\Delta^{-1} H_{{\rm F}\!,1}$, we have to inspect the kinetic constraints in Eq.~\eqref{eq:H-1}. In particular, the off-diagonal terms of $H_{{\rm F}\!,1}$ act nontrivially on configurations that contain any one of the following subsequences: $\downarrow\downarrow\downarrow$, $\downarrow\uparrow\downarrow$, $\downarrow\uparrow\downarrow\uparrow$, or $\uparrow\downarrow\uparrow\downarrow$ (some of them were frozen under $H^{(1)}$). Combined with the first-order frozen-state condition, which prohibits pairs of $\downarrow$-spins, we therefore have that $\mathcal{C}_2$ contains configurations without subsequences $\downarrow\downarrow$ or $\downarrow\uparrow\downarrow$.

The second correction, entering the truncation $H^{(3)}=H^{(2)}+\Delta^{-2}H_{{\rm F}\!,2}$, is reported in Eq.~\eqref{eq:H-2}. Inspecting the off-diagonal terms in that equation---in particular the second one---leads us to the conclusion that the configurations in $\mathcal{C}_3$ have no subsequences $\downarrow\downarrow$, $\downarrow\uparrow\downarrow$, or $\downarrow\uparrow\uparrow\downarrow$. The first two subsequences are forbidden since the state should be frozen under $H^{(2)}$, whereas the third subsequence is prohibited since the state should additionally be frozen under $H_{{\rm F}\!,2}$. 

More generally, the above observations suggest that, apart from the trivial vacuum state $\ket{\downarrow\downarrow\cdots\downarrow}$,
\begin{quote}
    \emph{$\mathcal{C}_k$ contains spin configurations in which any two $\downarrow$-spins are separated by at least $k$ $\uparrow$-spins.}
\end{quote}
In Appendix~\ref{app:counting} we show that the number of such configurations is\footnote{This formula is derived under the assumption that the system size $L$ is larger than the number of adjacent sites acted upon by the local terms in $H^{(k)}$, i.e., $L>2k+1$.}
\begin{align}
    |\mathcal{C}_k|\!=\!1\!+\!\!\sum_{N=\lceil\!\frac{kL}{k+1}\!\rceil}^L \!\!\frac{L}{kN\!-\!(k\!-\!1)L}{kN-(k\!-\!1)L\choose L\!-\!N}.
    \label{eq:number-frozen-states}
\end{align}
It correctly reproduces the numerically computed numbers of frozen states, shown in Fig.~\hyperref[fig:Krylov_stuff]{3(a)}. In Appendix~\ref{app:counting} we also derive the large-$L$ asymptotics of $|\mathcal{C}_k|$, obtaining
\begin{align}
    |\mathcal{C}_k|\sim\chi_k^L,
    \label{eq:asymptotics}
\end{align}
where $\chi_k$ is a solution of $\chi^k(\chi-1)=1$. For $k=1$ we have $\chi_1=(1+\sqrt{5})/2\approx 1.618$, and for $k=2$ the solution reads $\chi_2\approx 1.466$, both values matching the results previously obtained in Ref.~\cite{zadnik-fagotti-2021}. The numerics confirms the scaling also for higher-order truncations---see Fig.~\hyperref[fig:Krylov_stuff]{3(b)}. 
\begin{figure}
    \centering
    \includegraphics[width=1\linewidth]{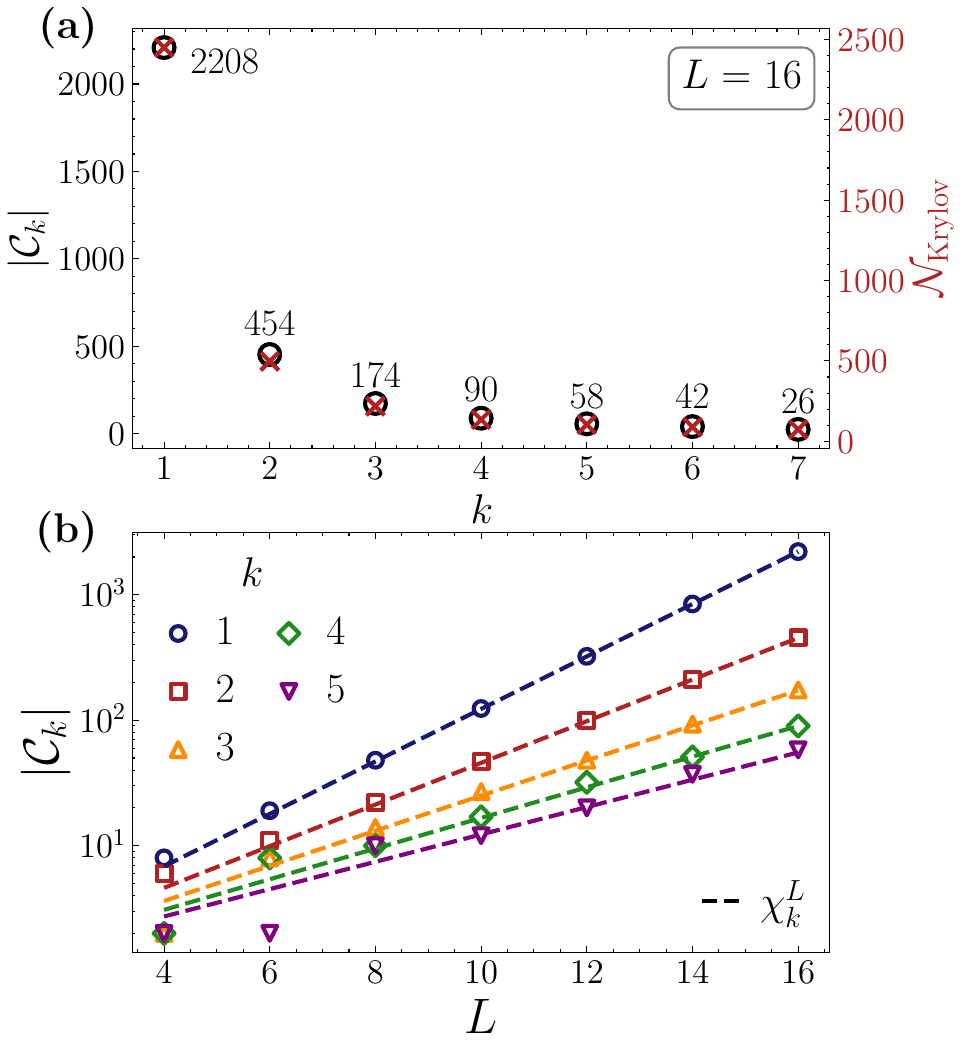}
    \caption{{\bf Frozen states and scaling of Krylov sectors.} Panel (a) shows the scaling of the number of frozen states $|\mathcal{C}_k|$ (black) and the number of Krylov sectors $\mathcal{N}_{\mathrm{Krylov}}$ (red) with order $k$ of the truncation, for system size $L = 16$. The values of $|\mathcal{C}_k|$ computed from Eq.~\eqref{eq:number-frozen-states} match the numerically computed ones shown in the plot. Panel (b) depicts the scaling of $|\mathcal{C}_k|$ with system size $L$ for various $k$. The dashed lines show the asymptotic prediction from Eq.~\eqref{eq:asymptotics}. In both panels the coupling was set to $\Delta = 4$.}
    \label{fig:Krylov_stuff}
\end{figure}

\begin{figure*}[t!]
    \centering
    \includegraphics[width=0.75\textwidth]{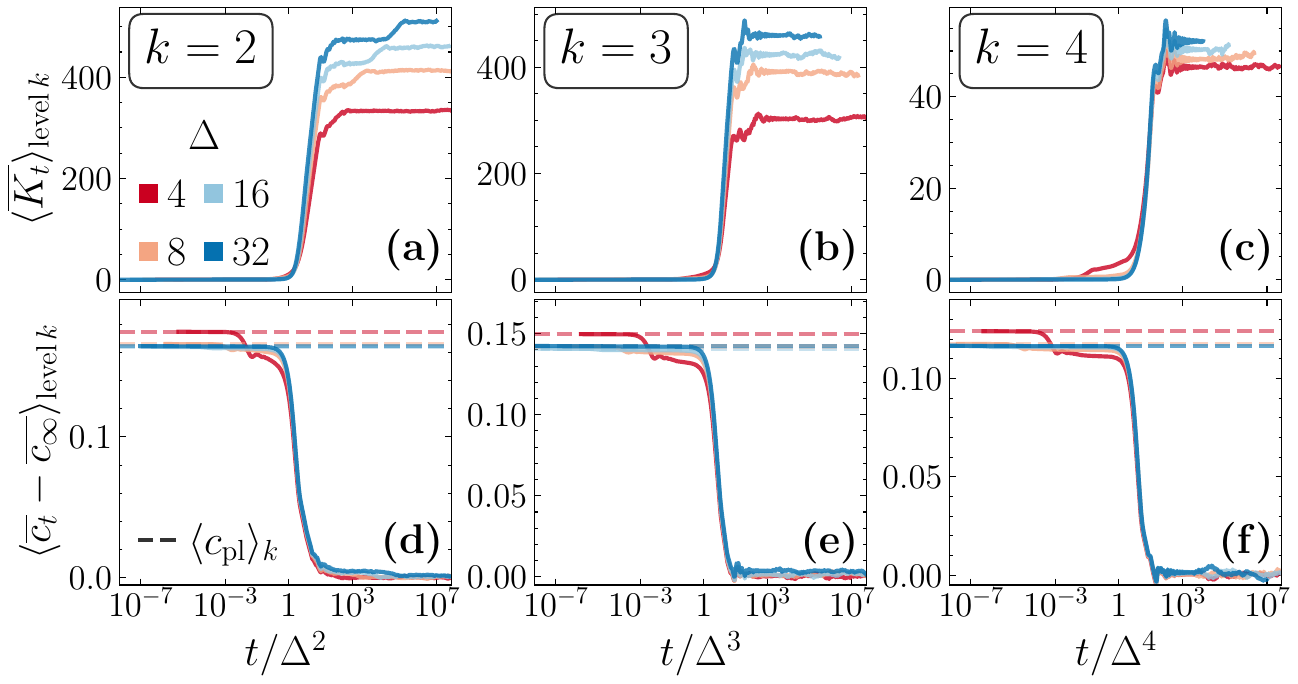}
    \caption{{\bf Krylov complexity and correlation functions in frozen states.} Panels (a)--(c) show time-averaged Krylov complexity for a level-$k$ initial state, additionally averaged over all initial states in the level. Notice the collapse in the rescaled time $t/\Delta^k$. Panels (d)--(f) show time-averaged correlation functions in level-$k$ states, averaged across the level. The dashed lines show the plateau-height estimates given in Eq.~\eqref{eq:plateau-value}. The  $O(\Delta^{-2})$ correction in Eq.~\eqref{eq:plateau-value} suggests that the two-staged plateaus (for $\Delta=4$) become single-staged as $\Delta\to\infty$. In all plots the system size is $L=14$.}
    \label{fig:krylov-complexity}
\end{figure*}

We remark that the integrability of $H_{{\rm F}\!,0}$ ensures the existence of stable quasiparticles~\cite{zadnik-fagotti-2021,pozsgay-etal-2021}. In frozen states belonging to $\mathcal{C}_1$ (apart from the trivial vacuum state), these quasiparticles are densely packed and immobile: the frozen states of $H_{{\rm F}\!,0}$ are genuinely {\em jammed}~\cite{bidzhiev-fagotti-zadnik-2022,zadnik-bocini-bidzhiev-fagotti-2022}. The leading order of the folded model $H_{\rm F}$ (and of the effective model $H$ as well) moreover exhibits Hilbert space fragmentation, in which the dynamically disconnected (Krylov) sectors are labeled by specific jammed configurations~\cite{zadnik-fagotti-2021,pozsgay-etal-2021,yang-etal-2020}. While higher orders of the folded model are no longer integrable, they still seem to exhibit Hilbert space fragmentation, although to a lesser degree. Furthermore, we observe a correlation between the number of Krylov sectors and the number of frozen states in higher-order truncations of $H$---see Fig.~\hyperref[fig:Krylov_stuff]{3(a)}: the scalings are the same up to a prefactor. This hints at a possible connection between the frozen states and the Hilbert space fragmentation also in higher orders of the effective Hamiltonian $H$. We leave this for future investigations. In the following we will instead demonstrate that the nested hierarchy of frozen states in Eq.~\eqref{eq:nested-hierarchy} gives rise to a hierarchy of time scales observed in dynamical correlation functions.

\section{Hierarchy of time scales} 

Expanding the effective Hamiltonian in the time-evolution operator as $\exp(-\ii t H)=\exp(-\ii t \Delta V-\ii \sum_{n\ge 0}t \Delta^{-n} H_{{\rm F}\!,n})$
suggests that there exists a hierarchy of time scales at which successive orders of the folded model $H_{\rm F}$ become relevant for the dynamics. These time scales correspond to the powers of the coupling strength $\Delta$. For example, until $t\sim \Delta^k$, at which the prefactor $t \Delta^{-k}$ in front of $H_{{\rm F}\!,k}$ becomes appreciable, we expect the dynamics to be governed by the truncation $H^{(k)}$ defined in Eq.~\eqref{eq:truncations}. This truncation has a set of frozen states that is distinct from those of the higher-order truncations. In particular, some of them are frozen for $H^{(k)}$ but not for $H^{(k+1)}$: they belong to the set $\mathcal{C}_k\setminus\mathcal{C}_{k+1}$, which we refer to as \emph{level $k$}---see Fig.~\ref{fig:hierarchy}. During the effective time evolution with $H$, we expect the states in level $k$ not to evolve up to times proportional to $\Delta^k$. At least for large $\Delta$, the same can be expected from the evolution under $H_{\rm XPX}$, related to $H$ via a unitary transformation $\ee^{S}$---see Eq.~\eqref{eq:large-coupling-expansion}. Indeed, since the leading order in $S$ is $1/\Delta$, we expect the unitary transformation to be irrelevant at large $\Delta$.

To verify our expectations, we compute how the complexity of level-$k$ frozen states evolves in time under the Hamiltonian $H_{\rm XPX}$, for various $k$ (see Appendix~\ref{app:FA-model} for similar results in the quantum Fredrickson-Andersen model). In particular, we consider the {\em Krylov} (or {\em spread}) {\em complexity} which measures the spreading of states through the Hilbert space during time evolution~\cite{balasubramanian-etal-2022,nandy-etal-2025}. For any initial state $\ket{\psi}$ it is defined as
\begin{align}
    K_t(\psi)=\sum_{n=0,1,2,\ldots} n\, |\!\braket{\psi(t) |B_n }\!|^2,
\end{align}
where $\ket{\psi(t)}=e^{-\ii t H_{\rm XPX}}\ket{\psi}$ is the time-evolved state and $\{\ket{B_n}\}_{n=0,1,2,\ldots}$ is the Krylov basis obtained by orthogonalizing the set of states $\{\ket{\psi_n}:=(H_{\rm XPX})^n\ket{\psi}\}_{n=0,1,2,\ldots}$.

In Figs.~\hyperref[fig:krylov-complexity]{4(a),\,(b),\,(c)} we show the time-averaged Krylov complexity $\overline{K_t}(\boldsymbol{s}):=t^{-1}\int _0^t \dd \tau K_\tau(\boldsymbol{s})$, additionally averaged over the frozen initial states $\ket{\psi}=\ket{\boldsymbol{s}}$ in different levels of the hierarchy. In particular, the complexity of states in level $k$ remains negligible up to times proportional to $\Delta^k$, after which it exhibits a jump. This is confirmed by the collapse of functions $\overline{K_t}(\boldsymbol{s})$ in the rescaled time $t/\Delta^k$. Interestingly, we observe that the unitary transformation $\ee^{S}$ has little significance even for intermediate values of $\Delta$.

Figures~\hyperref[fig:krylov-complexity]{4(d),\,(e),\,(f)} demonstrate that the hierarchy of time scales also appears in the correlation functions. While the time-averaged correlations $\overline{c_t}(\boldsymbol{s})-\overline{c_\infty}(\boldsymbol{s})$ rapidly decay to zero in nonfrozen states $\ket{\boldsymbol{s}}$ (see Fig.~\ref{fig:slow-relaxation-xpx}), those in level-$k$ frozen states exhibit plateaus that persist up to times proportional to $\Delta^k$. The estimate for the plateau value of $\overline{c_t}(\boldsymbol{s})$, derived in Appendix~\ref{app:plateau-values}, reads
\begin{align}
     c_{\rm pl}(\boldsymbol{s})=\frac{N_\uparrow(\boldsymbol{s})}{L}+O(\Delta^{-2}),
     \label{eq:plateau-value}
\end{align}
where $N_\uparrow(\boldsymbol{s})$ is the number of $\uparrow$-spins in the level-$k$ state $\ket{\boldsymbol{s}}$. As shown in Figs.~\hyperref[fig:krylov-complexity]{4(d),\,(e),\,(f)}, the difference $c_{\rm pl}(\boldsymbol{s})-\overline{c_\infty}(\boldsymbol{s})$ correctly reproduces the heights of the plateaus. Note that $c_{\rm pl}(\boldsymbol{s})$ coincides with the initial value of $\overline{c_t}(\boldsymbol{s})$ up to an $O(\Delta^{-2})$ correction. The latter arises from the unitary transformation $\ee^S$, since the state is evolved with $\ee^{-\ii t H_{\rm XPX}}=\ee^{-S}\ee^{-\ii t H}\ee^{S}$, not with $\ee^{-\ii t H}$. Note that some of the plateaus at intermediate values of $\Delta$, shown in Figs.~\hyperref[fig:krylov-complexity]{4(d),\,(e),\,(f)}, are not one- but two-staged on the associated time scale. The $O(\Delta^{-2})$ correction suggests that the second stages in such plateaus disappear as $\Delta\to\infty$ (plateaus become steps). The above results fully explain the hierarchy of relaxation times depicted in Fig.~\hyperref[fig:slow-relaxation-xpx]{1(a)}.

In general, the decay of correlation functions in level-$k$ frozen states can exhibit a richer structure than in the XPX model discussed above. In particular, it may proceed across multiple time scales $t \sim \Delta^{\ell_j}$, with $\ell_1 < \ell_2 < \cdots$ and starting with $\ell_1 = k$. We indeed observe such behavior in the quantum Fredrickson-Andersen model, as reported in Appendix~\ref{app:FA-model}. This suggests that, during the time evolution, components of the evolving superposition can reenter the nested hierarchy of frozen states, thereby giving rise to additional, longer relaxation time scales.

\section{Discussion} 

We have traced the emergence of slow relaxation in quantum kinetically constrained models to a nested hierarchy of computational-basis states that remain frozen up to times proportional to the powers of the coupling strength $\Delta$. For the concrete example of the XPX model, our findings are consistent with the observation of spatial heterogeneity in the dynamics, reported in Ref.~\cite{zadnik-garrahan-2023}. Indeed, the times up to which the states in the nested hierarchy remain frozen scale as $\Delta^k$, where $k$ is the smallest spatial separation between spins $\downarrow$. The latter facilitate the dynamics under the kinetic constraint. We can then expect the regions of space in which $\downarrow$-spins are further apart to relax exponentially slower than the regions where $\downarrow$-spins are closer. Crucially, in our case the exponential dependence of time scales on the spatial separation emerges in a clean (unperturbed) system, regardless of its integrability. The underlying mechanism is implicit: it is hidden in the large-coupling expansion, whose higher-order terms in $1/\Delta$ act on progressively larger clusters of spins. In contrast, a broad range of relaxation times can also arise from an explicit perturbation which breaks integrability or constrains the dynamics~\cite{michailidis-etal-2018,lisiecki-etal-2025}.

Via duality transformations, our results can explain the onset of slow relaxation also in quantum many-body systems in which kinetic constraints are not explicit, but emerge in the large-coupling limit~\cite{zadnik-garrahan-2023,causer-banuls-garrahan-2024}. Whether they can be applied to systems in which kinetic constraints are hard to pinpoint even in the large-coupling limit remains an intriguing open question~\cite{carleo-etal-2012,honda-etal-2025}. Another among open questions concerns the relation between the nested hierarchy of frozen states and emergent (quasi)conserved quantities~\cite{abanin-deroeck-ho-huveneers-2017,chen-iadecola-2021}. Last but not least, it would be interesting to see whether the nested hierarchy can be extended to include recently discovered Hilbert-space cages~\cite{tan-huang-2025,benami-heyl-moessner-2025,nicolau-ljubotina-serbyn-2025,jonay-pollmann-2025}, in addition to frozen states.

\smallskip
\noindent
{\bf \em Acknowledgements.} L.P.~is grateful to Urban Duh, Yusuf Kasim, and Pavel Orlov for useful discussions. ED calculations were performed using the QuSpin package \cite{QuSpin, QuSpin2}. This work has been supported by the Slovenian Research and Innovation Agency (ARIS) under Research Startup Program No. SN-ZRD/22-27/0510---NESY (L.Z. and V.M.) and Research Program No. P1-0402 (L.Z. and L.P.), as well as by the European Research Council (ERC) under Advanced Grant No. 101096208---QUEST (L.Z. and L.P.).

\appendix

\section{Large-coupling expansion}
\label{app:large-coupling}

Here, we report the details for the large-coupling expansion, largely based on Ref.~\cite{macdonald-1988}. The starting point is the separation of the Hamiltonian $H_{\rm XPX}$ as
\begin{align}
    H_{\rm XPX}=H_{\rm F\!,0}+T_{\!-\!1}+T_{1}+\Delta V.
    \label{eq:Hamiltonian-separation}
\end{align}
Here, $V=\sum_{j=1}^L \sigma_j^z$ is the interaction, whose spectrum is equally spaced. Operators $T_m$ satisfy $[V,T_m]=4 m T_m$ and $T_{-m}\equiv T_m^\dagger$. They cause transitions between the eigenvalues of $V$ that differ by integer multiples of $4$. Finally, $H_{\rm F\!,0}$ is chosen so as to commute with $V$, i.e., $[V,H_{\rm F\!,0}]=0$. In the limit $\Delta\to\infty$, the transitions between the eigenvalues of $\Delta V$, separated by a multiple of $\Delta$, are  prohibitively costly. The leading order of the large coupling expansion is therefore $H_{\rm F\!,0}+\Delta V$.

Let us introduce compact notation $T_0\equiv H_{\rm F\!,0}$ and $T_{m_1,\ldots, m_k}\equiv T_{m_1}\cdots T_{m_k}$, with $m_j\in\{-1,0,1\}$. In this paper we focus on the XPX model, for which $T_0$ is given in Eq.~\eqref{eq:H-0} in the main text, and
\begin{align} 
    T_1=\sum_{j=1}^L\sigma_{j-1}^+(\mathds{1}-\sigma_{j}^z)\sigma_{j+1}^+,
\end{align} 
with $\sigma_j^\pm=(\sigma^x_j\pm\ii\sigma^y_{j})/2$. We note, however,
that the following discussion works for any Hamiltonian with a separation analogous to Eq.~\eqref{eq:Hamiltonian-separation}.

The large-coupling expansion consists of anti-Hermitian operators $S^{(n)}$ and effective Hamiltonians $H^{(n)}$, for $n=1,2,\ldots\,$, such that
\begin{align}\label{expansion asymptotic}
    \ee^{S^{(n)}} H_{\rm XPX} \ee^{-S^{(n)}}=H^{(n)}+O(\Delta^{-n})
\end{align}
as $\Delta\to\infty$, and the effective Hamiltonians $H^{(n)}$ conserve $V$, i.e., $[H^{(n)},V]=0$. Consistently with numerical observations, we will regard the expansion obtained as $n\to\infty$ not only as asymptotic in $\Delta$, but as a convergent power series. Before proceeding, note that
\begin{align}
    [V,T_{m_1,\ldots, m_k}]=\left(4\sum_{j=1}^k m_j \right) T_{m_1,\ldots, m_k}.
\end{align}
Thus, if $\sum_{j=1}^k m_j=0$, the commutator is zero. In particular, the effective Hamiltonians will be linear combinations of terms $T_{m_1,\ldots,m_k}$ with $\sum_{j=1}^k m_j=0$, while $S^{(k)}$ will consist of such terms with $\sum_{j=1}^k m_j\ne 0$.

The recursive scheme for the large-coupling expansion is now as follows~\cite{macdonald-1988}:
\begin{enumerate}[leftmargin=*]
    \item We start with $S^{(1)}=(4\Delta)^{-1}(T_1-T_{\!-\!1})$.
    \item From
    \begin{align}
    \ee^{S^{(k)}} H_{\rm XPX} &\ee^{-S^{(k)}}=H^{(k)}\notag\\
    &+\Delta^{-k}\!\!\sum_{\substack{m_1\ldots m_{k+\!1}\\ \in\{\text{-}1,0,1\}}} \!\!C_{m_1,\ldots,m_{k+\!1}}T_{m_1,\ldots,m_{k+\!1}}\notag\\
    &+O(\Delta^{-(k+\!1)})  
    \end{align}
we obtain  $H^{(k)}$, consisting of terms of orders $\Delta,\Delta^0,\ldots, \Delta^{1-k}$, as well as the coefficients $C_{m_1,\ldots,m_{k+1}}$ describing the correction of order $\Delta^{-k}$.
    \item We construct
\begin{align}
    &S^{(k+\!1)}=S^{(k)}\notag\\
    &\hspace{1em}+\Delta^{-(k+\!1)}\!\!\sum_{\substack{m_1\ldots m_{k+1}\\ \sum_{j=1}^{k+1} m_j \neq 0 }}\!\!\tfrac{C_{m_1,\ldots,m_{k+1}}}{4\sum_{j=1}^{k+1} m_j}T_{m_1,\ldots,m_{k+1}},    
\end{align}
and repeat the steps recursively. In particular, this yields
\begin{align}
    H^{(k+\!1)}\!=&H^{(k)}\notag\\
    +&\Delta^{-k}\!\sum_{\substack{m_1\ldots m_{k+\!1}\\  \sum_{j=1}^{k+1} m_j = 0 }}\!\! C_{m_1,\ldots,m_{k+\!1}}T_{m_1,\ldots,m_{k+\!1}}.    
\end{align}
\end{enumerate}
For obtaining the coefficients in the second step of the recurrence, it is convenient to use the Campbell identity
\begin{align}
    \ee^{S^{(k)}} H_{\rm XPX} \ee^{-S^{(k)}}\!\!=\!\!\sum_{\ell=0}^{k+1}\!\frac{({\rm ad}_{S^{(k)}})^\ell}{\ell!} H_{\rm XPX}\!
    +\!O(\Delta^{-k-\!1}),
\end{align}
where ${\rm ad}_X Y:=[X,Y]$.

We will write the effective Hamiltonians (i.e., truncations of the expansion in $1/\Delta$) in the form $H^{(k)}=\Delta V + \sum_{n=0}^{k-1} \Delta^{-n} H_{{\rm F}\!,n}$ and the generator of the unitary transformation as $S^{(k)}=\sum_{n=1}^k\Delta^{-n}S_{n}$---see Eq.~\eqref{eq:truncations} in the main text. In this notation, $H_{\rm F}\equiv \sum_{n=0}^{\infty} \Delta^{-n} H_{{\rm F}\!,n}$ is the folded model. Different orders of the expansion are obtained using symbolic programming. The lowest orders (aside from $H_{\rm F\!,0}\equiv T_0$) read
\begin{align}
    H_{\rm F\!,1}\!=&\tfrac{1}{4}\big[T_{1,\!-\!1}\!-\!T_{\!-\!1, 1}\big],\notag\\
    H_{\rm F\!,2}\!=&\tfrac{1}{4^2}\big[T_{\!-\!1, 0, 1}\!-\!\tfrac{1}{2}T_{\!-\!1, 1, 0}\!-\!\tfrac{1}{2}T_{0,\!-\!1, 1}\!-\!\tfrac{1}{2}T_{0, 1, \!-\!1}\notag\\
    &-\!\tfrac{1}{2}T_{1,\!-\!1, 0}\!+\!T_{1, 0, \!-\!1}\big], \notag\\
    H_{\rm F\!,3}\!=&\tfrac{1}{4^3}\big[\!-\!\tfrac{1}{2}T_{\!-\!1,\!-\!1, 1, 1}\!-\!T_{\!-\!1, 0, 0, 1}\!+\!T_{\!-\!1, 0, 1, 0}\!+\!T_{\!-\!1, 1, \!-\!1,1} \notag\\
    &-\!\tfrac{1}{2}T_{\!-\!1, 1, 0, 0}\!+\!T_{0,\!-\!1, 0, 1}\!-\!\tfrac{1}{2}T_{0, 0,\!-\!1, 1}\!+\!\tfrac{1}{2}T_{0, 0, 1,\!-\!1} \notag\\
    &-\!T_{0, 1, 0,\!-\!1}\!+\!\tfrac{1}{2}T_{1, \!-\!1, 0, 0}\!-\!T_{1,\!-\!1, 1, \!-\!1}\!-\!T_{1, 0,\!-\!1, 0}\notag\\
    &+\!T_{1, 0, 0,\!-\!1}\!+\!\tfrac{1}{2}T_{1, 1,\!-\!1,\!-\!1}\big],
    \label{eq:folded-orders}
\end{align}
and
\begin{align}
    S_1\!=&\tfrac{1}{4}\big[T_1\!-\!T_{\!-\!1}\big],\notag\\
    S_2\!=&\tfrac{1}{4^2}[T_{\!-\!1, 0}\!-\!T_{0,\!-\!1}\!-\!T_{0,1}\!+\!T_{1, 0}\big],\notag\\
    S_3\!=&\tfrac{1}{4^3}\big[\tfrac{1}{4}T_{\!-\!1,\!-\!1,0}\!-\!\tfrac{2}{3}T_{\!-\!1,\!-\!1, 1}\!-\!\tfrac{1}{2}T_{\!-\!1, 0, \!-\!1}\!-\!T_{\!-\!1, 0, 0}\notag\\ 
    &+\!\tfrac{4}{3}T_{\!-\!1, 1,\!-\!1}\!+\!\tfrac{2}{3}T_{\!-\!1, 1, 1}\!+\!\tfrac{1}{4}T_{0,\!-\!1,\!-\!1}\!+\!2T_{0,\!-\!1,0}\!-\!\tfrac{1}{4}T_{1,1,0}\notag\\ 
    &-\!T_{0, 0,\!-\!1}\!+\!T_{0, 0, 1}\!-\!2T_{0, 1, 0}\!-\!\tfrac{1}{4}T_{0, 1, 1}\!-\!\tfrac{2}{3}T_{1,\!-\!1,\!-\!1}\notag\\ 
    &+\!T_{1, 0, 0}\!-\!\tfrac{4}{3}T_{1,\!-\!1,1}\!+\!\tfrac{1}{2}T_{1, 0, 1}\!+\!\tfrac{2}{3}T_{1,1,\!-\!1}\big].
    \label{eq:S-orders}
\end{align}

\begin{figure}[h!]
    \centering    \includegraphics[width=\linewidth]{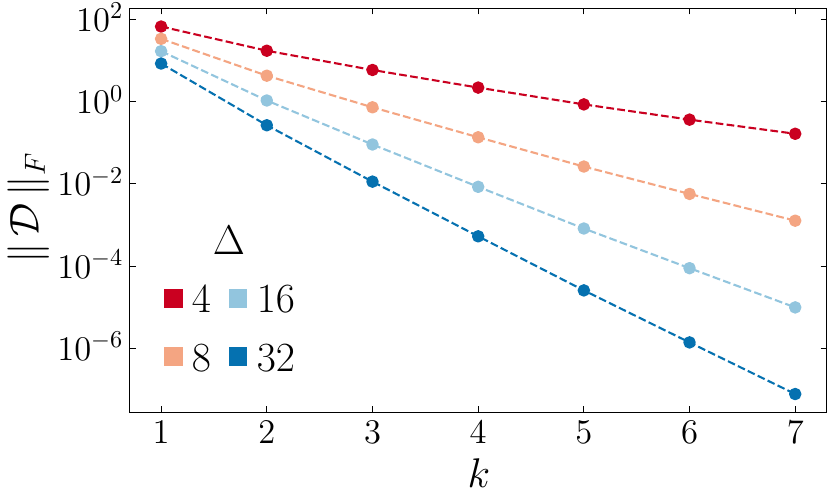}
   \caption{{\bf Convergence of the large-coupling expansion}. Frobenius norm of the difference $\mathcal{D} = H_{\rm XPX} - e^{-S^{(k)}} H^{(k)}  e^{S^{(k)}}$ as a function of the truncation order $k$, for different $\Delta>1$. System size is $L=12$.}
    \label{fig:convergence}
\end{figure}

By construction, the introduced large-coupling expansion is asymptotic as $\Delta\to\infty$ [see Eq.~\eqref{expansion asymptotic}]. However, it seems also to be convergent in $k$ for fixed $\Delta>1$. Some evidence is presented in Fig.~\ref{fig:convergence}.

\section{Counting the frozen states}
\label{app:counting}

Here, we compute the size of the set $\mathcal{C}_k$, which contains the frozen states of $H^{(k)}$. We assume that the system size $L$ is larger than the number of neighboring sites on which the local terms in $H^{(k)}$ act: $L>2k+1$. According to our conjecture, any two $\downarrow$-spins in a state that belongs to $\mathcal{C}_k$ should be separated by at least $k$ spins $\uparrow$. We start by counting such states at a fixed total number $N$ of $\uparrow$-spins ($L-N$ is then the number of $\downarrow$-spins). 

\smallskip
\noindent
{\bf \em Step 1.} First consider the blocks of spins 
\begin{align}
    A:=\,\downarrow\underbrace{\uparrow\ldots\uparrow}_{k},\qquad
    B:=\,\uparrow.
\end{align}
We can arrange spins into $N_A:=L-N$ blocks $A$ and $N_B:=N-k(L-N)$ blocks $B$. From here, we can already deduce the possible values of the total number of $\uparrow$-spins, $N$: since the minimal number of $B$-s is $N^{\rm (min)}_B=0$, we have $\lceil kL/(k+1)\rceil \le N\le L$. We can concatenate $A$-s and $B$-s in
\begin{align}
    {N_A+N_B\choose N_A}={L-k(L-N)\choose L-N}
    \label{eq:step-1}
\end{align}
different ways. Because of the way we arranged spins into the block $A$, the configurations obtained in this way end in $k$ or more $\uparrow$-spins on the right-hand side (RHS). We now have to count the rest of configurations---those that end in less than $k$ $\uparrow$-spins on the RHS.

\smallskip
\noindent
{\bf \em Step 2.} Let us fix the rightmost block to be $A$, remove $\ell\in\{1,\ldots,k\}$ of the rightmost $\uparrow$-spins in it, and move them to the left-hand side. We obtain
\begin{align}
    \underbrace{\uparrow\ldots\uparrow}_{\ell}\,\Bigl|\qquad\cdots A\cdots B\cdots\qquad\,\Bigr|\downarrow\underbrace{\uparrow\ldots\uparrow}_{k-\ell},
\end{align}  
where we have to concatenate $N_A-1$ remaining blocks $A$ and $N_B$ blocks $B$ in the middle. There are 
\begin{align}
    {N_A-1+N_B\choose N_A-1}={L-k(L-N)-1\choose L-N-1}
\end{align}
such concatenations. Since $\ell$ can run from $1$ to $k$, we therefore gain
\begin{align}
    k{L-k(L-N)-1\choose L-N-1}
    \label{eq:step-2}
\end{align}
additional configurations in this way. The total number of frozen states in $\mathcal{C}_k$ that have $N$ $\uparrow$-spins is now the sum of contributions in Eqs.~\eqref{eq:step-1} and~\eqref{eq:step-2}:
\begin{align}
    {L-k(L-N)\choose L-N}+k{L-k(L-N)-1\choose L-N-1}=\notag\\
    =\frac{L}{kN-(k-1)L}{kN-(k-1)L\choose L-N}.
    \label{eq:fixed-N-frozen}
\end{align}

The above results generalize (and agree with) the results obtained in Ref.~\cite{zadnik-fagotti-2021}---see Eqs.~(56) and~(64) therein. Summing over all possible values of $N$, and adding $1$ to account also for the trivial frozen state $\ket{\downarrow\downarrow\cdots\downarrow}$, we obtain $|\mathcal{C}_k|$, reported in Eq.~\eqref{eq:number-frozen-states} in the main text.

Let us now assume that the asymptotic contribution to Eq.~\eqref{eq:number-frozen-states} in the main text comes from the dominant term in the sum. The latter has the form given in Eq.~\eqref{eq:fixed-N-frozen}, for some $N$ which we will find by maximization. We expect $|\mathcal{C}_k|$ to scale exponentially with $L$ and consider the logarithm of Eq.~\eqref{eq:fixed-N-frozen}, which has two terms,
\begin{align}
    \log\frac{L}{kN\!-\!(k\!-\!1)L}+\log{kN-(k\!-\!1)L\choose L\!-\!N}.
\end{align}
The second one scales linearly with $L$, while the first one represents a logarithmic correction which we will neglect. We then set $N=\rho L$, use Stirling's approximation for $\log{L[1-k(1-\rho)]\choose L(1-\rho)}$, and maximize with respect to $\rho$. This leads to
\begin{align}
    |\mathcal{C}_k|\!\sim\! \left\{\!\!\frac{[1\!-\!k(1\!-\!\rho_k)]^{1-k(1-\rho_k)}}{(1\!-\!\rho_k)^{1-\rho_k}[1\!-\!(k\!+\!1)(1\!-\!\rho_k)]^{1-(k+1)(1-\rho_k)}}\!\!\right\}^L\!\!,
    \label{eq:asymptotics-1}
\end{align}
where $\rho_k$ is the solution of
\begin{align}
    (1\!-\!\rho)\frac{[1\!-\!k(1\!-\!\rho)]^k}{[1\!-\!(k\!+\!1)(1\!-\!\rho)]^{k+1}}\!=\!1.
    \label{eq:asymptotics-2}
\end{align}
Using this constraint we can further simplify Eq.~\eqref{eq:asymptotics-1}. In particular, defining 
\begin{align}
    \chi_k:=\frac{1-k(1-\rho_k)}{1-(k+1)(1-\rho_k)},    
\end{align}
Eq.~\eqref{eq:asymptotics-1} becomes $|\mathcal{C}_k|\sim\chi_k^L$ [Eq.~\eqref{eq:asymptotics} in the main text] while Eq.~\eqref{eq:asymptotics-2} translates into the constraint $\chi^k(\chi-1)=1$ satisfied by $\chi_k$.

\section{Plateau values of the correlation functions}
\label{app:plateau-values}

Here, we estimate the plateau value of the correlation function $\overline{c_t}(\boldsymbol{s})$, evaluated in a level-$k$ frozen state, $\ket{\boldsymbol{s}}\in\mathcal{C}_k\setminus\mathcal{C}_{k+1}$. We first recall that the dynamics on time scales $t\sim\Delta^k$, on which the plateau appears, is governed by $\ee^{-S}\ee^{-\ii t H^{(k)}}\ee^{S}$. At larger times, higher orders of the effective Hamiltonian kick in and cause the eventual decay of the plateau. In their absence, the correlation function $\overline{c_t}(\boldsymbol{s})$ would retain its plateau value forever. The plateau height, denoted by $c_{\rm pl}(\boldsymbol{s})$, should therefore be equal to the $t\to\infty$ limit of the correlation function evolved under $\ee^{-S}\ee^{-\ii t H^{(k)}}\ee^{S}$. We have
\begin{align}
    c_{\rm pl}(\boldsymbol{s})\!=\!\sum_{j=1}^L\frac{\delta_{\uparrow,s_j}}{L}\lim_{t\to\infty}\!\frac{1}{t}\!\int_0^t \!\dd \tau\!\bra{\psi_{\boldsymbol{s}}(\tau)}\!\ee^{S} n_j \ee^{-S}\!\ket{\psi_{\boldsymbol{s}}(\tau)},
    \label{eq:plateau-1}
\end{align}
where we have defined $\ket{\psi_{\boldsymbol{s}}(\tau)}:=\ee^{-\ii \tau H^{(k)}}\ee^{S}\ket{\boldsymbol{s}}$ and used that $n_j$ is diagonal in the computational basis, $\braket{\boldsymbol{s}|n_j|\boldsymbol{s}}=\delta_{\uparrow,s_j}$. To simplify the time average in Eq.~\eqref{eq:plateau-1}, we now choose a basis $\{\ket{E_\ell}\}$, such that $H^{(k)}\ket{E_\ell}=E_\ell \ket{E_\ell}$ and $V\ket{E_\ell}=v_\ell \ket{E_\ell}$: this is possible since $[H^{(k)},V]=0$. We obtain
\begin{align}
    &\lim_{t\to\infty}\!\frac{1}{t}\!\int_0^t \!\!\dd \tau\!\bra{\psi_{\boldsymbol{s}}(\tau)}\!\ee^{S} n_j \ee^{-S}\!\ket{\psi_{\boldsymbol{s}}(\tau)}\!=\notag\\
    &=\!\sum_{E_\ell=E_m}\!\!\braket{\boldsymbol{s}|\ee^{-S}|E_\ell}\!\!\braket{E_\ell|\ee^S n_j \ee^{-S}|E_m}\!\!\braket{E_m|\ee^S|\boldsymbol{s}}\!=\notag\\
    &=\!\sum_{E_\ell=E_m}\!\!\braket{\boldsymbol{s}|E_\ell}\!\!\braket{E_\ell|n_j|E_m}\!\!\braket{E_m|\boldsymbol{s}}\!+\!O(\Delta^{-2}).
    \label{eq:time-average-part}
\end{align}
In passing to the third row, we have used the Campbell identity 
\begin{align}
    \ee^{-S}\!\ket{E_\ell}\!\!\bra{E_\ell}\!\ee^S\!=\!\ket{E_\ell}\!\!\bra{E_\ell}\!-\!\Delta^{-1}\!\big[S_1,\ket{E_\ell}\!\!\bra{E_\ell}\!\big]\!+\!O(\Delta^{-2}).
\end{align}
When plugged in Eq.~\eqref{eq:time-average-part}, the first order in $1/\Delta$ disappears, since $\bra{\boldsymbol{s}}\big[S_1,\ket{E_\ell}\!\!\bra{E_\ell}\big]n_j \ket{E_m}\!\!\braket{E_m |\boldsymbol{s}}=0$. This is because $n_j$ and $\ket{E_\ell}\!\!\bra{E_\ell}$ preserve the magnetization $V$, while $S_1$ changes it---see Eq.~\eqref{eq:S-1} in the main text. We now note that
\begin{align}
&\sum_{E_\ell=E_m}\!\!\braket{\boldsymbol{s}|E_\ell}\!\!\braket{E_\ell|n_j|E_m}\!\!\braket{E_m|\boldsymbol{s}}\!=\notag\\
    &=\!\lim_{t\to\infty}\!\frac{1}{t}\int_0^t\!\!\dd\tau\braket{\boldsymbol{s}|\ee^{\ii\tau H^{(k)}} n_j\ee^{-\ii\tau H^{(k)}}|\boldsymbol{s}}\!=\!\braket{\boldsymbol{s}|n_j|\boldsymbol{s}},
\end{align}
where we have used that $\ket{\boldsymbol{s}}\in\mathcal{C}_k\setminus\mathcal{C}_{k+1}$ remains frozen under $H^{(k)}$. Equation~\eqref{eq:plateau-1} thus simplifies into
\begin{align}
     c_{\rm pl}(\boldsymbol{s})\!=\!&\sum_{j=1}^L \frac{\delta_{\uparrow,s_j}}{L}\!+\!O(\Delta^{-2}),
\end{align}
which is Eq.~\eqref{eq:plateau-value} in the main text.\\

\section{Quantum Fredrickson-Andersen model}
\label{app:FA-model}

The quantum version of the Fredrickson-Andersen (FA) model~\cite{ritort-sollich-2003,fredrickson-andersen-1984} reads
\begin{equation}
    H_{\rm FA}\!=\!\sum_{j=1}^L n_{j-1} \!\left\{ \!\sqrt{c(1\!-\!c)} \sigma_j^x \!-\! {\rm e}^s \left[c(1\!-\!n_j)\!+\!(1\!-\!c)n_j\right] \right\},
\end{equation}
where $s\in\mathbb{R}$ is the potential energy strength and $c\in(0,1)$ is an additional parameter. We will restrict to the special point $c=2/3$, where the model becomes
\begin{equation}
    H_{\rm FA}\!=\!\frac{1}{3}\sum_{j=1}^L (n_{j-1}\!+\!n_{j+1})\left[\sqrt{2}\sigma_j^x\!-\!{\rm e}^s(2\!-\!n_j) \right].
\end{equation}
The stochastic point separating the active and inactive regimes is $s = \log\sqrt{2}$. 

The large-coupling expansion can be applied with no additional considerations apart for the starting point:
\begin{equation}
H_{\rm FA}=H_{\rm F\!,0}+T_{\!-\!1}+T_{1}+\Delta V,
\end{equation}
where
\begin{align} 
    & H_{{\rm F}\!,0}=\frac{2\sqrt{2}}{3}\sum_{\ell=1}^L n_{\ell-1}\sigma_\ell^x n_{\ell+1},\\
    & T_1=\frac{\sqrt{2}}{6}\sum_{\ell=1}^L (1\!-\!\sigma^z_{\ell-1}\sigma^z_{\ell+1})\sigma_\ell^-,\\
    & V=2\sum_{\ell=1}^L (n_{\ell-1}\!+\!n_{\ell+1})(n_\ell\!-\!2),
\end{align}
and we denote $\Delta={\rm e}^s/6$. Then, Eqs.~\eqref{eq:folded-orders} and~\eqref{eq:S-orders} can be used. Note that $H_{\rm{F}\!, 0}$ is the PXP model~\cite{lesanovsky-2012,turner-etal-2018}.

\begin{figure}[h!]
    \centering    \includegraphics[width=\linewidth]{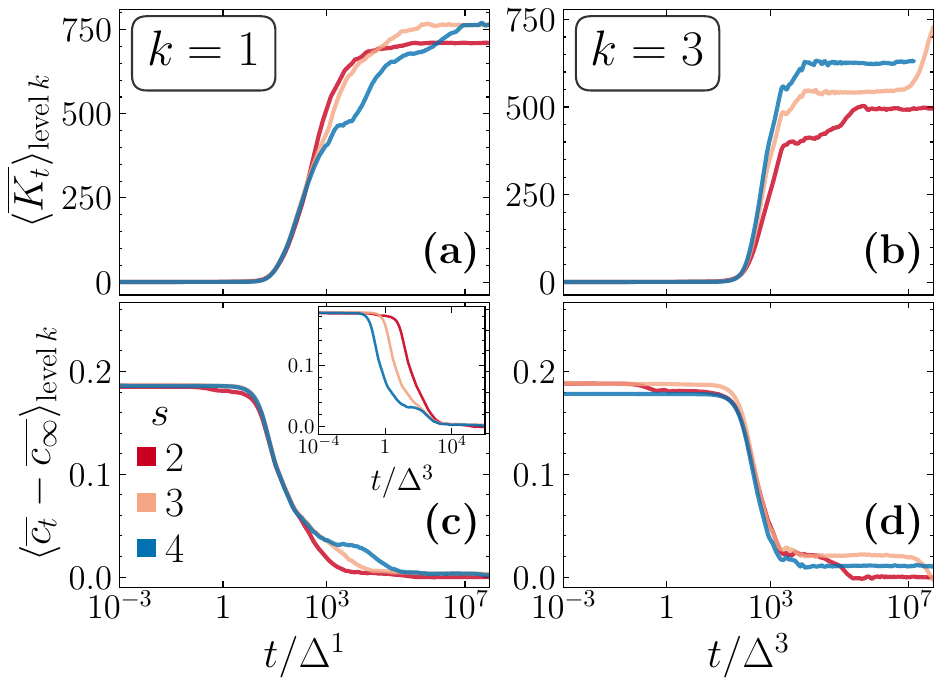}
   \caption{{\bf Krylov complexity and correlations in frozen states for the quantum FA model.} Panels (a) and (b) show time-averaged Krylov complexity for a level-$k$ initial state, for $k=1,3$, additionally averaged over all initial states in the level. Notice the collapse in the rescaled time $t/\Delta^k$, where $\Delta={\rm e}^s/6$. Panels (c) and (d) show time-averaged correlation functions $\overline{c_t}(\boldsymbol{s})=t^{-1} \int_0^t\dd \tau c_\tau(\boldsymbol{s})$, Eq.~\eqref{eq:correlation} in the main text, in level-$k$ states, averaged across the level. The states in level $k=1$ have secondary plateaus that collapse in the rescaled time $t/\Delta^3$, as shown in the inset of panel (c). This indicates that, for times larger than $t\sim \Delta$, such states evolve into a superposition which includes frozen states belonging to level $k=3$. Levels $k=2, 4$ in this model are empty, so we omit them. In both plots the system size is $L=12$.}
    \label{fig:krylovFA}
\end{figure}

\begin{figure}[h!]
    \centering    \includegraphics[width=\linewidth]{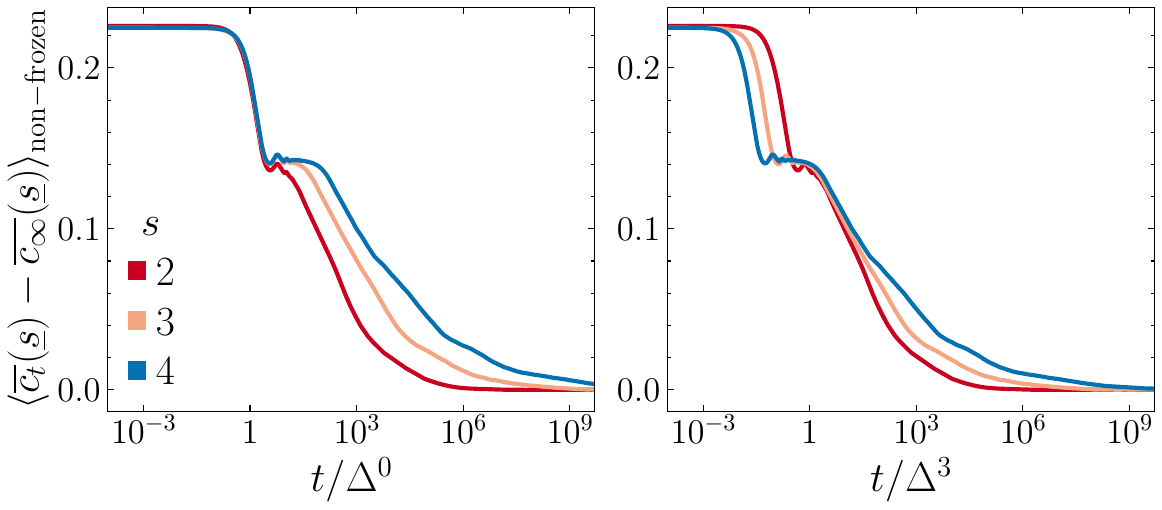}
   \caption{{\bf Correlations in nonfrozen states for the quantum FA model.} Correlation functions, averaged over all nonfrozen states, are plotted for various values of $\Delta=\ee^s/6$, in two different time scales: $t/\Delta^0$ and $t/\Delta^3$.}
    \label{fig:nonfrozenFA}
\end{figure}

In Fig.~\ref{fig:krylovFA} we present the time evolution of Krylov complexity and correlation functions for the lowest nonempty levels in the hierarchy of states. Similar to the XPX model discussed in the main text, we observe a collapse in the rescaled time $t/\Delta^k$. Contrary to the XPX model, however, the relaxation of correlation functions in the quantum FA model is multi-staged even for larger values of $\Delta$. For example, for initial states in the level $k=1$, we observe a secondary plateau with a characteristic time scale $t\sim\Delta^3$, as evident in the inset of Fig.~\hyperref[fig:krylovFA]{6(c)}. A similar two-staged relaxation also appears for correlations in nonfrozen states, as depicted in Fig.~\ref{fig:nonfrozenFA}. The time scales associated with different relaxation stages suggest that initial states $\ket{\psi}$ evolve into superpositions that include frozen states within higher levels of the nested hierarchy, which relax on longer time scales.

\bibliography{references.bib}

\begin{thebibliography}{58}%
\makeatletter
\providecommand \@ifxundefined [1]{%
 \@ifx{#1\undefined}
}%
\providecommand \@ifnum [1]{%
 \ifnum #1\expandafter \@firstoftwo
 \else \expandafter \@secondoftwo
 \fi
}%
\providecommand \@ifx [1]{%
 \ifx #1\expandafter \@firstoftwo
 \else \expandafter \@secondoftwo
 \fi
}%
\providecommand \natexlab [1]{#1}%
\providecommand \enquote  [1]{``#1''}%
\providecommand \bibnamefont  [1]{#1}%
\providecommand \bibfnamefont [1]{#1}%
\providecommand \citenamefont [1]{#1}%
\providecommand \href@noop [0]{\@secondoftwo}%
\providecommand \href [0]{\begingroup \@sanitize@url \@href}%
\providecommand \@href[1]{\@@startlink{#1}\@@href}%
\providecommand \@@href[1]{\endgroup#1\@@endlink}%
\providecommand \@sanitize@url [0]{\catcode `\\12\catcode `\$12\catcode
  `\&12\catcode `\#12\catcode `\^12\catcode `\_12\catcode `\%12\relax}%
\providecommand \@@startlink[1]{}%
\providecommand \@@endlink[0]{}%
\providecommand \url  [0]{\begingroup\@sanitize@url \@url }%
\providecommand \@url [1]{\endgroup\@href {#1}{\urlprefix }}%
\providecommand \urlprefix  [0]{URL }%
\providecommand \Eprint [0]{\href }%
\providecommand \doibase [0]{https://doi.org/}%
\providecommand \selectlanguage [0]{\@gobble}%
\providecommand \bibinfo  [0]{\@secondoftwo}%
\providecommand \bibfield  [0]{\@secondoftwo}%
\providecommand \translation [1]{[#1]}%
\providecommand \BibitemOpen [0]{}%
\providecommand \bibitemStop [0]{}%
\providecommand \bibitemNoStop [0]{.\EOS\space}%
\providecommand \EOS [0]{\spacefactor3000\relax}%
\providecommand \BibitemShut  [1]{\csname bibitem#1\endcsname}%
\let\auto@bib@innerbib\@empty
\bibitem [{\citenamefont {Palmer}\ \emph {et~al.}(1984)\citenamefont {Palmer},
  \citenamefont {Stein}, \citenamefont {Abrahams},\ and\ \citenamefont
  {Anderson}}]{palmer-etal-1984}%
  \BibitemOpen
  \bibfield  {author} {\bibinfo {author} {\bibfnamefont {R.~G.}\ \bibnamefont
  {Palmer}}, \bibinfo {author} {\bibfnamefont {D.~L.}\ \bibnamefont {Stein}},
  \bibinfo {author} {\bibfnamefont {E.}~\bibnamefont {Abrahams}},\ and\
  \bibinfo {author} {\bibfnamefont {P.~W.}\ \bibnamefont {Anderson}},\
  }\bibfield  {title} {\bibinfo {title} {{Models of hierarchically constrained
  dynamics for glassy relaxation}},\ }\href
  {https://doi.org/10.1103/PhysRevLett.53.958} {\bibfield  {journal} {\bibinfo
  {journal} {Phys. Rev. Lett.}\ }\textbf {\bibinfo {volume} {53}},\ \bibinfo
  {pages} {958} (\bibinfo {year} {1984})}\BibitemShut {NoStop}%
\bibitem [{\citenamefont {Fredrickson}\ and\ \citenamefont
  {Andersen}(1984)}]{fredrickson-andersen-1984}%
  \BibitemOpen
  \bibfield  {author} {\bibinfo {author} {\bibfnamefont {G.~H.}\ \bibnamefont
  {Fredrickson}}\ and\ \bibinfo {author} {\bibfnamefont {H.~C.}\ \bibnamefont
  {Andersen}},\ }\bibfield  {title} {\bibinfo {title} {{Kinetic Ising model of
  the glass transition}},\ }\href {https://doi.org/10.1103/PhysRevLett.53.1244}
  {\bibfield  {journal} {\bibinfo  {journal} {Phys. Rev. Lett.}\ }\textbf
  {\bibinfo {volume} {53}},\ \bibinfo {pages} {1244} (\bibinfo {year}
  {1984})}\BibitemShut {NoStop}%
\bibitem [{\citenamefont {Ritort}\ and\ \citenamefont
  {Sollich}(2003)}]{ritort-sollich-2003}%
  \BibitemOpen
  \bibfield  {author} {\bibinfo {author} {\bibfnamefont {F.}~\bibnamefont
  {Ritort}}\ and\ \bibinfo {author} {\bibfnamefont {P.}~\bibnamefont
  {Sollich}},\ }\bibfield  {title} {\bibinfo {title} {{Glassy dynamics of
  kinetically constrained models}},\ }\href
  {https://doi.org/10.1080/0001873031000093582} {\bibfield  {journal} {\bibinfo
   {journal} {Advances in Physics}\ }\textbf {\bibinfo {volume} {52}},\
  \bibinfo {pages} {219} (\bibinfo {year} {2003})}\BibitemShut {NoStop}%
\bibitem [{\citenamefont {Garrahan}(2018)}]{garrahan-2018}%
  \BibitemOpen
  \bibfield  {author} {\bibinfo {author} {\bibfnamefont {J.~P.}\ \bibnamefont
  {Garrahan}},\ }\bibfield  {title} {\bibinfo {title} {{Aspects of
  non-equilibrium in classical and quantum systems: Slow relaxation and
  glasses, dynamical large deviations, quantum non-ergodicity, and open quantum
  dynamics}},\ }\href
  {https://doi.org/https://doi.org/10.1016/j.physa.2017.12.149} {\bibfield
  {journal} {\bibinfo  {journal} {Physica A}\ }\textbf {\bibinfo {volume}
  {504}},\ \bibinfo {pages} {130} (\bibinfo {year} {2018})}\BibitemShut
  {NoStop}%
\bibitem [{\citenamefont {Urban}\ \emph {et~al.}(2009)\citenamefont {Urban},
  \citenamefont {Johnson}, \citenamefont {Henage}, \citenamefont {Isenhower},
  \citenamefont {Yavuz}, \citenamefont {Walker},\ and\ \citenamefont
  {Saffman}}]{urban-etal-2009}%
  \BibitemOpen
  \bibfield  {author} {\bibinfo {author} {\bibfnamefont {E.}~\bibnamefont
  {Urban}}, \bibinfo {author} {\bibfnamefont {T.~A.}\ \bibnamefont {Johnson}},
  \bibinfo {author} {\bibfnamefont {T.}~\bibnamefont {Henage}}, \bibinfo
  {author} {\bibfnamefont {L.}~\bibnamefont {Isenhower}}, \bibinfo {author}
  {\bibfnamefont {D.~D.}\ \bibnamefont {Yavuz}}, \bibinfo {author}
  {\bibfnamefont {T.~G.}\ \bibnamefont {Walker}},\ and\ \bibinfo {author}
  {\bibfnamefont {M.}~\bibnamefont {Saffman}},\ }\bibfield  {title} {\bibinfo
  {title} {{Observation of Rydberg blockade between two atoms}},\ }\href
  {https://doi.org/10.1038/nphys1178} {\bibfield  {journal} {\bibinfo
  {journal} {Nature Physics}\ }\textbf {\bibinfo {volume} {5}},\ \bibinfo
  {pages} {110–114} (\bibinfo {year} {2009})}\BibitemShut {NoStop}%
\bibitem [{\citenamefont {Lesanovsky}(2011)}]{lesanovsky-2011}%
  \BibitemOpen
  \bibfield  {author} {\bibinfo {author} {\bibfnamefont {I.}~\bibnamefont
  {Lesanovsky}},\ }\bibfield  {title} {\bibinfo {title} {{Many-body spin
  interactions and the ground state of a dense Rydberg lattice gas}},\ }\href
  {https://doi.org/10.1103/PhysRevLett.106.025301} {\bibfield  {journal}
  {\bibinfo  {journal} {Phys. Rev. Lett.}\ }\textbf {\bibinfo {volume} {106}},\
  \bibinfo {pages} {025301} (\bibinfo {year} {2011})}\BibitemShut {NoStop}%
\bibitem [{\citenamefont {Lesanovsky}\ and\ \citenamefont
  {Garrahan}(2013)}]{lesanovsky-garrahan-2013}%
  \BibitemOpen
  \bibfield  {author} {\bibinfo {author} {\bibfnamefont {I.}~\bibnamefont
  {Lesanovsky}}\ and\ \bibinfo {author} {\bibfnamefont {J.~P.}\ \bibnamefont
  {Garrahan}},\ }\bibfield  {title} {\bibinfo {title} {{Kinetic constraints,
  hierarchical relaxation, and onset of glassiness in strongly interacting and
  dissipative Rydberg gases}},\ }\href
  {https://doi.org/10.1103/PhysRevLett.111.215305} {\bibfield  {journal}
  {\bibinfo  {journal} {Phys. Rev. Lett.}\ }\textbf {\bibinfo {volume} {111}},\
  \bibinfo {pages} {215305} (\bibinfo {year} {2013})}\BibitemShut {NoStop}%
\bibitem [{\citenamefont {Bernien}\ \emph {et~al.}(2017)\citenamefont
  {Bernien}, \citenamefont {Schwartz}, \citenamefont {Keesling}, \citenamefont
  {Levine}, \citenamefont {Omran}, \citenamefont {Pichler}, \citenamefont
  {Choi}, \citenamefont {Zibrov}, \citenamefont {Endres}, \citenamefont
  {Greiner}, \citenamefont {Vuleti{\'{c}}},\ and\ \citenamefont
  {Lukin}}]{bernien-etal-2017}%
  \BibitemOpen
  \bibfield  {author} {\bibinfo {author} {\bibfnamefont {H.}~\bibnamefont
  {Bernien}}, \bibinfo {author} {\bibfnamefont {S.}~\bibnamefont {Schwartz}},
  \bibinfo {author} {\bibfnamefont {A.}~\bibnamefont {Keesling}}, \bibinfo
  {author} {\bibfnamefont {H.}~\bibnamefont {Levine}}, \bibinfo {author}
  {\bibfnamefont {A.}~\bibnamefont {Omran}}, \bibinfo {author} {\bibfnamefont
  {H.}~\bibnamefont {Pichler}}, \bibinfo {author} {\bibfnamefont
  {S.}~\bibnamefont {Choi}}, \bibinfo {author} {\bibfnamefont {A.~S.}\
  \bibnamefont {Zibrov}}, \bibinfo {author} {\bibfnamefont {M.}~\bibnamefont
  {Endres}}, \bibinfo {author} {\bibfnamefont {M.}~\bibnamefont {Greiner}},
  \bibinfo {author} {\bibfnamefont {V.}~\bibnamefont {Vuleti{\'{c}}}},\ and\
  \bibinfo {author} {\bibfnamefont {M.~D.}\ \bibnamefont {Lukin}},\ }\bibfield
  {title} {\bibinfo {title} {{Probing many-body dynamics on a 51-atom quantum
  simulator}},\ }\href {https://doi.org/10.1038/nature24622} {\bibfield
  {journal} {\bibinfo  {journal} {Nature}\ }\textbf {\bibinfo {volume} {551}},\
  \bibinfo {pages} {579} (\bibinfo {year} {2017})}\BibitemShut {NoStop}%
\bibitem [{\citenamefont {Bluvstein}\ \emph {et~al.}(2021)\citenamefont
  {Bluvstein}, \citenamefont {Omran}, \citenamefont {Levine}, \citenamefont
  {Keesling}, \citenamefont {Semeghini}, \citenamefont {Ebadi}, \citenamefont
  {Wang}, \citenamefont {Michailidis}, \citenamefont {Maskara}, \citenamefont
  {Ho}, \citenamefont {Choi}, \citenamefont {Serbyn}, \citenamefont {Greiner},
  \citenamefont {Vuletić},\ and\ \citenamefont {Lukin}}]{bluvstein-etal-2021}%
  \BibitemOpen
  \bibfield  {author} {\bibinfo {author} {\bibfnamefont {D.}~\bibnamefont
  {Bluvstein}}, \bibinfo {author} {\bibfnamefont {A.}~\bibnamefont {Omran}},
  \bibinfo {author} {\bibfnamefont {H.}~\bibnamefont {Levine}}, \bibinfo
  {author} {\bibfnamefont {A.}~\bibnamefont {Keesling}}, \bibinfo {author}
  {\bibfnamefont {G.}~\bibnamefont {Semeghini}}, \bibinfo {author}
  {\bibfnamefont {S.}~\bibnamefont {Ebadi}}, \bibinfo {author} {\bibfnamefont
  {T.~T.}\ \bibnamefont {Wang}}, \bibinfo {author} {\bibfnamefont {A.~A.}\
  \bibnamefont {Michailidis}}, \bibinfo {author} {\bibfnamefont
  {N.}~\bibnamefont {Maskara}}, \bibinfo {author} {\bibfnamefont {W.~W.}\
  \bibnamefont {Ho}}, \bibinfo {author} {\bibfnamefont {S.}~\bibnamefont
  {Choi}}, \bibinfo {author} {\bibfnamefont {M.}~\bibnamefont {Serbyn}},
  \bibinfo {author} {\bibfnamefont {M.}~\bibnamefont {Greiner}}, \bibinfo
  {author} {\bibfnamefont {V.}~\bibnamefont {Vuletić}},\ and\ \bibinfo
  {author} {\bibfnamefont {M.~D.}\ \bibnamefont {Lukin}},\ }\bibfield  {title}
  {\bibinfo {title} {{Controlling quantum many-body dynamics in driven Rydberg
  atom arrays}},\ }\href {https://doi.org/10.1126/science.abg2530} {\bibfield
  {journal} {\bibinfo  {journal} {Science}\ }\textbf {\bibinfo {volume}
  {371}},\ \bibinfo {pages} {1355} (\bibinfo {year} {2021})}\BibitemShut
  {NoStop}%
\bibitem [{\citenamefont {Kim}\ \emph {et~al.}(2024)\citenamefont {Kim},
  \citenamefont {Yang}, \citenamefont {M{\o{}}lmer},\ and\ \citenamefont
  {Ahn}}]{kim-etal-2024}%
  \BibitemOpen
  \bibfield  {author} {\bibinfo {author} {\bibfnamefont {K.}~\bibnamefont
  {Kim}}, \bibinfo {author} {\bibfnamefont {F.}~\bibnamefont {Yang}}, \bibinfo
  {author} {\bibfnamefont {K.}~\bibnamefont {M{\o{}}lmer}},\ and\ \bibinfo
  {author} {\bibfnamefont {J.}~\bibnamefont {Ahn}},\ }\bibfield  {title}
  {\bibinfo {title} {{Realization of an Extremely Anisotropic Heisenberg Magnet
  in Rydberg Atom Arrays}},\ }\href
  {https://doi.org/10.1103/PhysRevX.14.011025} {\bibfield  {journal} {\bibinfo
  {journal} {Phys. Rev. X}\ }\textbf {\bibinfo {volume} {14}},\ \bibinfo
  {pages} {011025} (\bibinfo {year} {2024})}\BibitemShut {NoStop}%
\bibitem [{\citenamefont {Yang}\ \emph {et~al.}(2025)\citenamefont {Yang},
  \citenamefont {Yarloo}, \citenamefont {Zhang}, \citenamefont {M{\o{}}lmer},\
  and\ \citenamefont {Nielsen}}]{yang-etal-2025}%
  \BibitemOpen
  \bibfield  {author} {\bibinfo {author} {\bibfnamefont {F.}~\bibnamefont
  {Yang}}, \bibinfo {author} {\bibfnamefont {H.}~\bibnamefont {Yarloo}},
  \bibinfo {author} {\bibfnamefont {H.-C.}\ \bibnamefont {Zhang}}, \bibinfo
  {author} {\bibfnamefont {K.}~\bibnamefont {M{\o{}}lmer}},\ and\ \bibinfo
  {author} {\bibfnamefont {A.~E.~B.}\ \bibnamefont {Nielsen}},\ }\bibfield
  {title} {\bibinfo {title} {{Probing Hilbert space fragmentation with strongly
  interacting Rydberg atoms}},\ }\href
  {https://doi.org/10.1103/PhysRevB.111.144313} {\bibfield  {journal} {\bibinfo
   {journal} {Phys. Rev. B}\ }\textbf {\bibinfo {volume} {111}},\ \bibinfo
  {pages} {144313} (\bibinfo {year} {2025})}\BibitemShut {NoStop}%
\bibitem [{\citenamefont {Corcoran}\ \emph {et~al.}(2025)\citenamefont
  {Corcoran}, \citenamefont {de~Leeuw},\ and\ \citenamefont
  {Pozsgay}}]{corcoran-etal-2025}%
  \BibitemOpen
  \bibfield  {author} {\bibinfo {author} {\bibfnamefont {L.}~\bibnamefont
  {Corcoran}}, \bibinfo {author} {\bibfnamefont {M.}~\bibnamefont {de~Leeuw}},\
  and\ \bibinfo {author} {\bibfnamefont {B.}~\bibnamefont {Pozsgay}},\
  }\bibfield  {title} {\bibinfo {title} {{Integrable models on Rydberg atom
  chains}},\ }\href {https://doi.org/10.21468/SciPostPhys.18.4.139} {\bibfield
  {journal} {\bibinfo  {journal} {SciPost Phys.}\ }\textbf {\bibinfo {volume}
  {18}},\ \bibinfo {pages} {139} (\bibinfo {year} {2025})}\BibitemShut
  {NoStop}%
\bibitem [{\citenamefont {Yang}\ \emph {et~al.}(2020)\citenamefont {Yang},
  \citenamefont {Liu}, \citenamefont {Gorshkov},\ and\ \citenamefont
  {Iadecola}}]{yang-etal-2020}%
  \BibitemOpen
  \bibfield  {author} {\bibinfo {author} {\bibfnamefont {Z.-C.}\ \bibnamefont
  {Yang}}, \bibinfo {author} {\bibfnamefont {F.}~\bibnamefont {Liu}}, \bibinfo
  {author} {\bibfnamefont {A.~V.}\ \bibnamefont {Gorshkov}},\ and\ \bibinfo
  {author} {\bibfnamefont {T.}~\bibnamefont {Iadecola}},\ }\bibfield  {title}
  {\bibinfo {title} {{Hilbert-space fragmentation from strict confinement}},\
  }\href {https://doi.org/10.1103/PhysRevLett.124.207602} {\bibfield  {journal}
  {\bibinfo  {journal} {Phys. Rev. Lett.}\ }\textbf {\bibinfo {volume} {124}},\
  \bibinfo {pages} {207602} (\bibinfo {year} {2020})}\BibitemShut {NoStop}%
\bibitem [{\citenamefont {Langlett}\ and\ \citenamefont
  {Xu}(2021)}]{langlett-xu-2021}%
  \BibitemOpen
  \bibfield  {author} {\bibinfo {author} {\bibfnamefont {C.~M.}\ \bibnamefont
  {Langlett}}\ and\ \bibinfo {author} {\bibfnamefont {S.}~\bibnamefont {Xu}},\
  }\bibfield  {title} {\bibinfo {title} {{Hilbert space fragmentation and exact
  scars of generalized Fredkin spin chains}},\ }\href
  {https://doi.org/10.1103/PhysRevB.103.L220304} {\bibfield  {journal}
  {\bibinfo  {journal} {Phys. Rev. B}\ }\textbf {\bibinfo {volume} {103}},\
  \bibinfo {pages} {L220304} (\bibinfo {year} {2021})}\BibitemShut {NoStop}%
\bibitem [{\citenamefont {Zadnik}\ and\ \citenamefont
  {Fagotti}(2021)}]{zadnik-fagotti-2021}%
  \BibitemOpen
  \bibfield  {author} {\bibinfo {author} {\bibfnamefont {L.}~\bibnamefont
  {Zadnik}}\ and\ \bibinfo {author} {\bibfnamefont {M.}~\bibnamefont
  {Fagotti}},\ }\bibfield  {title} {\bibinfo {title} {{The folded spin-1/2 XXZ
  model: I. Diagonalisation, jamming, and ground state properties}},\ }\href
  {https://doi.org/10.21468/SciPostPhysCore.4.2.010} {\bibfield  {journal}
  {\bibinfo  {journal} {SciPost Phys. Core}\ }\textbf {\bibinfo {volume} {4}},\
  \bibinfo {pages} {010} (\bibinfo {year} {2021})}\BibitemShut {NoStop}%
\bibitem [{\citenamefont {Pozsgay}\ \emph {et~al.}(2021)\citenamefont
  {Pozsgay}, \citenamefont {Gombor}, \citenamefont {Hutsalyuk}, \citenamefont
  {Jiang}, \citenamefont {Pristy{\'a}k},\ and\ \citenamefont
  {Vernier}}]{pozsgay-etal-2021}%
  \BibitemOpen
  \bibfield  {author} {\bibinfo {author} {\bibfnamefont {B.}~\bibnamefont
  {Pozsgay}}, \bibinfo {author} {\bibfnamefont {T.}~\bibnamefont {Gombor}},
  \bibinfo {author} {\bibfnamefont {A.}~\bibnamefont {Hutsalyuk}}, \bibinfo
  {author} {\bibfnamefont {Y.}~\bibnamefont {Jiang}}, \bibinfo {author}
  {\bibfnamefont {L.}~\bibnamefont {Pristy{\'a}k}},\ and\ \bibinfo {author}
  {\bibfnamefont {E.}~\bibnamefont {Vernier}},\ }\bibfield  {title} {\bibinfo
  {title} {{Integrable spin chain with Hilbert space fragmentation and solvable
  real-time dynamics}},\ }\href {https://doi.org/10.1103/PhysRevE.104.044106}
  {\bibfield  {journal} {\bibinfo  {journal} {Phys. Rev. E}\ }\textbf {\bibinfo
  {volume} {104}},\ \bibinfo {pages} {044106} (\bibinfo {year}
  {2021})}\BibitemShut {NoStop}%
\bibitem [{\citenamefont {Tamura}\ and\ \citenamefont
  {Katsura}(2022)}]{tamura-katsura-2022}%
  \BibitemOpen
  \bibfield  {author} {\bibinfo {author} {\bibfnamefont {K.}~\bibnamefont
  {Tamura}}\ and\ \bibinfo {author} {\bibfnamefont {H.}~\bibnamefont
  {Katsura}},\ }\bibfield  {title} {\bibinfo {title} {{Quantum many-body scars
  of spinless fermions with density-assisted hopping in higher dimensions}},\
  }\href {https://doi.org/10.1103/PhysRevB.106.144306} {\bibfield  {journal}
  {\bibinfo  {journal} {Phys. Rev. B}\ }\textbf {\bibinfo {volume} {106}},\
  \bibinfo {pages} {144306} (\bibinfo {year} {2022})}\BibitemShut {NoStop}%
\bibitem [{\citenamefont {Kerschbaumer}\ \emph {et~al.}(2025)\citenamefont
  {Kerschbaumer}, \citenamefont {Ljubotina}, \citenamefont {Serbyn},\ and\
  \citenamefont {Desaules}}]{kerschbaumer-etal-2025}%
  \BibitemOpen
  \bibfield  {author} {\bibinfo {author} {\bibfnamefont {A.}~\bibnamefont
  {Kerschbaumer}}, \bibinfo {author} {\bibfnamefont {M.}~\bibnamefont
  {Ljubotina}}, \bibinfo {author} {\bibfnamefont {M.}~\bibnamefont {Serbyn}},\
  and\ \bibinfo {author} {\bibfnamefont {J.-Y.}\ \bibnamefont {Desaules}},\
  }\bibfield  {title} {\bibinfo {title} {{Quantum many-body scars beyond the
  PXP model in Rydberg simulators}},\ }\href
  {https://doi.org/10.1103/PhysRevLett.134.160401} {\bibfield  {journal}
  {\bibinfo  {journal} {Phys. Rev. Lett.}\ }\textbf {\bibinfo {volume} {134}},\
  \bibinfo {pages} {160401} (\bibinfo {year} {2025})}\BibitemShut {NoStop}%
\bibitem [{\citenamefont {Moudgalya}\ \emph {et~al.}(2022)\citenamefont
  {Moudgalya}, \citenamefont {Bernevig},\ and\ \citenamefont
  {Regnault}}]{moudgalya-bernevig-ragnault-2022}%
  \BibitemOpen
  \bibfield  {author} {\bibinfo {author} {\bibfnamefont {S.}~\bibnamefont
  {Moudgalya}}, \bibinfo {author} {\bibfnamefont {B.~A.}\ \bibnamefont
  {Bernevig}},\ and\ \bibinfo {author} {\bibfnamefont {N.}~\bibnamefont
  {Regnault}},\ }\bibfield  {title} {\bibinfo {title} {{Quantum many-body scars
  and Hilbert space fragmentation: a review of exact results}},\ }\href
  {https://doi.org/10.1088/1361-6633/ac73a0} {\bibfield  {journal} {\bibinfo
  {journal} {Reports on Progress in Physics}\ }\textbf {\bibinfo {volume}
  {85}},\ \bibinfo {pages} {086501} (\bibinfo {year} {2022})}\BibitemShut
  {NoStop}%
\bibitem [{\citenamefont {Serbyn}\ \emph {et~al.}(2021)\citenamefont {Serbyn},
  \citenamefont {Abanin},\ and\ \citenamefont
  {Papi{\'{c}}}}]{serbyn-abanin-papic-2021}%
  \BibitemOpen
  \bibfield  {author} {\bibinfo {author} {\bibfnamefont {M.}~\bibnamefont
  {Serbyn}}, \bibinfo {author} {\bibfnamefont {D.~A.}\ \bibnamefont {Abanin}},\
  and\ \bibinfo {author} {\bibfnamefont {Z.}~\bibnamefont {Papi{\'{c}}}},\
  }\bibfield  {title} {\bibinfo {title} {{Quantum many-body scars and weak
  breaking of ergodicity}},\ }\href
  {https://doi.org/10.1038/s41567-021-01230-2} {\bibfield  {journal} {\bibinfo
  {journal} {Nature Physics}\ }\textbf {\bibinfo {volume} {17}},\ \bibinfo
  {pages} {675} (\bibinfo {year} {2021})}\BibitemShut {NoStop}%
\bibitem [{\citenamefont {Surace}\ \emph {et~al.}(2021)\citenamefont {Surace},
  \citenamefont {Votto}, \citenamefont {Lazo}, \citenamefont {Silva},
  \citenamefont {Dalmonte},\ and\ \citenamefont {Giudici}}]{surace-2021}%
  \BibitemOpen
  \bibfield  {author} {\bibinfo {author} {\bibfnamefont {F.~M.}\ \bibnamefont
  {Surace}}, \bibinfo {author} {\bibfnamefont {M.}~\bibnamefont {Votto}},
  \bibinfo {author} {\bibfnamefont {E.~G.}\ \bibnamefont {Lazo}}, \bibinfo
  {author} {\bibfnamefont {A.}~\bibnamefont {Silva}}, \bibinfo {author}
  {\bibfnamefont {M.}~\bibnamefont {Dalmonte}},\ and\ \bibinfo {author}
  {\bibfnamefont {G.}~\bibnamefont {Giudici}},\ }\bibfield  {title} {\bibinfo
  {title} {{Exact many-body scars and their stability in constrained quantum
  chains}},\ }\href {https://doi.org/10.1103/PhysRevB.103.104302} {\bibfield
  {journal} {\bibinfo  {journal} {Phys. Rev. B}\ }\textbf {\bibinfo {volume}
  {103}},\ \bibinfo {pages} {104302} (\bibinfo {year} {2021})}\BibitemShut
  {NoStop}%
\bibitem [{\citenamefont {Singh}\ \emph {et~al.}(2021)\citenamefont {Singh},
  \citenamefont {Ware}, \citenamefont {Vasseur},\ and\ \citenamefont
  {Friedman}}]{singh-etal-2021}%
  \BibitemOpen
  \bibfield  {author} {\bibinfo {author} {\bibfnamefont {H.}~\bibnamefont
  {Singh}}, \bibinfo {author} {\bibfnamefont {B.~A.}\ \bibnamefont {Ware}},
  \bibinfo {author} {\bibfnamefont {R.}~\bibnamefont {Vasseur}},\ and\ \bibinfo
  {author} {\bibfnamefont {A.~J.}\ \bibnamefont {Friedman}},\ }\bibfield
  {title} {\bibinfo {title} {{Subdiffusion and many-body quantum chaos with
  kinetic constraints}},\ }\href
  {https://doi.org/10.1103/PhysRevLett.127.230602} {\bibfield  {journal}
  {\bibinfo  {journal} {Phys. Rev. Lett.}\ }\textbf {\bibinfo {volume} {127}},\
  \bibinfo {pages} {230602} (\bibinfo {year} {2021})}\BibitemShut {NoStop}%
\bibitem [{\citenamefont {Yang}(2022)}]{yang-2022}%
  \BibitemOpen
  \bibfield  {author} {\bibinfo {author} {\bibfnamefont {Z.-C.}\ \bibnamefont
  {Yang}},\ }\bibfield  {title} {\bibinfo {title} {{Distinction between
  transport and R\'enyi entropy growth in kinetically constrained models}},\
  }\href {https://doi.org/10.1103/PhysRevB.106.L220303} {\bibfield  {journal}
  {\bibinfo  {journal} {Phys. Rev. B}\ }\textbf {\bibinfo {volume} {106}},\
  \bibinfo {pages} {L220303} (\bibinfo {year} {2022})}\BibitemShut {NoStop}%
\bibitem [{\citenamefont {McCarthy}\ \emph {et~al.}(2025)\citenamefont
  {McCarthy}, \citenamefont {Singh}, \citenamefont {Gopalakrishnan},\ and\
  \citenamefont {Vasseur}}]{mccarthy-etal-2025}%
  \BibitemOpen
  \bibfield  {author} {\bibinfo {author} {\bibfnamefont {C.}~\bibnamefont
  {McCarthy}}, \bibinfo {author} {\bibfnamefont {H.}~\bibnamefont {Singh}},
  \bibinfo {author} {\bibfnamefont {S.}~\bibnamefont {Gopalakrishnan}},\ and\
  \bibinfo {author} {\bibfnamefont {R.}~\bibnamefont {Vasseur}},\ }\bibfield
  {title} {\bibinfo {title} {{Subdiffusive transport in the Fredkin dynamical
  universality class}},\ }\href {https://doi.org/10.1103/PhysRevB.111.184317}
  {\bibfield  {journal} {\bibinfo  {journal} {Phys. Rev. B}\ }\textbf {\bibinfo
  {volume} {111}},\ \bibinfo {pages} {184317} (\bibinfo {year}
  {2025})}\BibitemShut {NoStop}%
\bibitem [{\citenamefont {Bhakuni}\ \emph {et~al.}(2025)\citenamefont
  {Bhakuni}, \citenamefont {Verdel}, \citenamefont {Desaules}, \citenamefont
  {Serbyn}, \citenamefont {Ljubotina},\ and\ \citenamefont
  {Dalmonte}}]{bhakuni-etal-2025}%
  \BibitemOpen
  \bibfield  {author} {\bibinfo {author} {\bibfnamefont {D.~S.}\ \bibnamefont
  {Bhakuni}}, \bibinfo {author} {\bibfnamefont {R.}~\bibnamefont {Verdel}},
  \bibinfo {author} {\bibfnamefont {J.-Y.}\ \bibnamefont {Desaules}}, \bibinfo
  {author} {\bibfnamefont {M.}~\bibnamefont {Serbyn}}, \bibinfo {author}
  {\bibfnamefont {M.}~\bibnamefont {Ljubotina}},\ and\ \bibinfo {author}
  {\bibfnamefont {M.}~\bibnamefont {Dalmonte}},\ }\bibfield  {title} {\bibinfo
  {title} {{Anomalously fast transport in non-integrable lattice gauge
  theories}},\ }\Eprint {https://arxiv.org/abs/2509.08889} {arXiv:2509.08889
  [cond-mat.quant-gas]}  (\bibinfo {year} {2025})\BibitemShut {NoStop}%
\bibitem [{\citenamefont {van Horssen}\ \emph {et~al.}(2015)\citenamefont {van
  Horssen}, \citenamefont {Levi},\ and\ \citenamefont
  {Garrahan}}]{horssen-levi-garrahan-2015}%
  \BibitemOpen
  \bibfield  {author} {\bibinfo {author} {\bibfnamefont {M.}~\bibnamefont {van
  Horssen}}, \bibinfo {author} {\bibfnamefont {E.}~\bibnamefont {Levi}},\ and\
  \bibinfo {author} {\bibfnamefont {J.~P.}\ \bibnamefont {Garrahan}},\
  }\bibfield  {title} {\bibinfo {title} {{Dynamics of many-body localization in
  a translation-invariant quantum glass model}},\ }\href
  {https://doi.org/10.1103/PhysRevB.92.100305} {\bibfield  {journal} {\bibinfo
  {journal} {Phys. Rev. B}\ }\textbf {\bibinfo {volume} {92}},\ \bibinfo
  {pages} {100305} (\bibinfo {year} {2015})}\BibitemShut {NoStop}%
\bibitem [{\citenamefont {Lan}\ \emph {et~al.}(2018)\citenamefont {Lan},
  \citenamefont {van Horssen}, \citenamefont {Powell},\ and\ \citenamefont
  {Garrahan}}]{lan-horssen-powell-garrahan-2018}%
  \BibitemOpen
  \bibfield  {author} {\bibinfo {author} {\bibfnamefont {Z.}~\bibnamefont
  {Lan}}, \bibinfo {author} {\bibfnamefont {M.}~\bibnamefont {van Horssen}},
  \bibinfo {author} {\bibfnamefont {S.}~\bibnamefont {Powell}},\ and\ \bibinfo
  {author} {\bibfnamefont {J.~P.}\ \bibnamefont {Garrahan}},\ }\bibfield
  {title} {\bibinfo {title} {{Quantum slow relaxation and metastability due to
  dynamical constraints}},\ }\href
  {https://doi.org/10.1103/PhysRevLett.121.040603} {\bibfield  {journal}
  {\bibinfo  {journal} {Phys. Rev. Lett.}\ }\textbf {\bibinfo {volume} {121}},\
  \bibinfo {pages} {040603} (\bibinfo {year} {2018})}\BibitemShut {NoStop}%
\bibitem [{\citenamefont {Feldmeier}\ \emph {et~al.}(2019)\citenamefont
  {Feldmeier}, \citenamefont {Pollmann},\ and\ \citenamefont
  {Knap}}]{feldmeier-pollmann-knap-2019}%
  \BibitemOpen
  \bibfield  {author} {\bibinfo {author} {\bibfnamefont {J.}~\bibnamefont
  {Feldmeier}}, \bibinfo {author} {\bibfnamefont {F.}~\bibnamefont
  {Pollmann}},\ and\ \bibinfo {author} {\bibfnamefont {M.}~\bibnamefont
  {Knap}},\ }\bibfield  {title} {\bibinfo {title} {{Emergent glassy dynamics in
  a quantum dimer model}},\ }\href
  {https://doi.org/10.1103/PhysRevLett.123.040601} {\bibfield  {journal}
  {\bibinfo  {journal} {Phys. Rev. Lett.}\ }\textbf {\bibinfo {volume} {123}},\
  \bibinfo {pages} {040601} (\bibinfo {year} {2019})}\BibitemShut {NoStop}%
\bibitem [{\citenamefont {Roy}\ and\ \citenamefont
  {Lazarides}(2020)}]{roy-lazarides-2020}%
  \BibitemOpen
  \bibfield  {author} {\bibinfo {author} {\bibfnamefont {S.}~\bibnamefont
  {Roy}}\ and\ \bibinfo {author} {\bibfnamefont {A.}~\bibnamefont
  {Lazarides}},\ }\bibfield  {title} {\bibinfo {title} {{Strong ergodicity
  breaking due to local constraints in a quantum system}},\ }\href
  {https://doi.org/10.1103/PhysRevResearch.2.023159} {\bibfield  {journal}
  {\bibinfo  {journal} {Phys. Rev. Res.}\ }\textbf {\bibinfo {volume} {2}},\
  \bibinfo {pages} {023159} (\bibinfo {year} {2020})}\BibitemShut {NoStop}%
\bibitem [{\citenamefont {Pancotti}\ \emph {et~al.}(2020)\citenamefont
  {Pancotti}, \citenamefont {Giudice}, \citenamefont {Cirac}, \citenamefont
  {Garrahan},\ and\ \citenamefont
  {Ba{\~n}uls}}]{pancotti-giudice-cirac-garrahan-banuls-2020}%
  \BibitemOpen
  \bibfield  {author} {\bibinfo {author} {\bibfnamefont {N.}~\bibnamefont
  {Pancotti}}, \bibinfo {author} {\bibfnamefont {G.}~\bibnamefont {Giudice}},
  \bibinfo {author} {\bibfnamefont {J.~I.}\ \bibnamefont {Cirac}}, \bibinfo
  {author} {\bibfnamefont {J.~P.}\ \bibnamefont {Garrahan}},\ and\ \bibinfo
  {author} {\bibfnamefont {M.~C.}\ \bibnamefont {Ba{\~n}uls}},\ }\bibfield
  {title} {\bibinfo {title} {Quantum east model: Localization, nonthermal
  eigenstates, and slow dynamics},\ }\href
  {https://doi.org/10.1103/PhysRevX.10.021051} {\bibfield  {journal} {\bibinfo
  {journal} {Phys. Rev. X}\ }\textbf {\bibinfo {volume} {10}},\ \bibinfo
  {pages} {021051} (\bibinfo {year} {2020})}\BibitemShut {NoStop}%
\bibitem [{\citenamefont {Zadnik}\ and\ \citenamefont
  {Garrahan}(2023)}]{zadnik-garrahan-2023}%
  \BibitemOpen
  \bibfield  {author} {\bibinfo {author} {\bibfnamefont {L.}~\bibnamefont
  {Zadnik}}\ and\ \bibinfo {author} {\bibfnamefont {J.~P.}\ \bibnamefont
  {Garrahan}},\ }\bibfield  {title} {\bibinfo {title} {{Slow heterogeneous
  relaxation due to constraints in dual XXZ models}},\ }\href
  {https://doi.org/10.1103/PhysRevB.108.L100304} {\bibfield  {journal}
  {\bibinfo  {journal} {Phys. Rev. B}\ }\textbf {\bibinfo {volume} {108}},\
  \bibinfo {pages} {L100304} (\bibinfo {year} {2023})}\BibitemShut {NoStop}%
\bibitem [{\citenamefont {Causer}\ \emph {et~al.}(2024)\citenamefont {Causer},
  \citenamefont {Ba{\~n}uls},\ and\ \citenamefont
  {Garrahan}}]{causer-banuls-garrahan-2024}%
  \BibitemOpen
  \bibfield  {author} {\bibinfo {author} {\bibfnamefont {L.}~\bibnamefont
  {Causer}}, \bibinfo {author} {\bibfnamefont {M.~C.}\ \bibnamefont
  {Ba{\~n}uls}},\ and\ \bibinfo {author} {\bibfnamefont {J.~P.}\ \bibnamefont
  {Garrahan}},\ }\bibfield  {title} {\bibinfo {title} {{Nonthermal eigenstates
  and slow relaxation in quantum Fredkin spin chains}},\ }\href
  {https://doi.org/10.1103/PhysRevB.110.134322} {\bibfield  {journal} {\bibinfo
   {journal} {Phys. Rev. B}\ }\textbf {\bibinfo {volume} {110}},\ \bibinfo
  {pages} {134322} (\bibinfo {year} {2024})}\BibitemShut {NoStop}%
\bibitem [{\citenamefont {Menzler}\ \emph {et~al.}(2025)\citenamefont
  {Menzler}, \citenamefont {Ba{\~n}uls},\ and\ \citenamefont
  {Heidrich-Meisner}}]{menzler-banuls-meisner-2025}%
  \BibitemOpen
  \bibfield  {author} {\bibinfo {author} {\bibfnamefont {H.~G.}\ \bibnamefont
  {Menzler}}, \bibinfo {author} {\bibfnamefont {M.~C.}\ \bibnamefont
  {Ba{\~n}uls}},\ and\ \bibinfo {author} {\bibfnamefont {F.}~\bibnamefont
  {Heidrich-Meisner}},\ }\bibfield  {title} {\bibinfo {title} {{Graph theory
  and tunable slow dynamics in quantum East Hamiltonians}},\ }\href
  {https://doi.org/10.1103/j7jf-746f} {\bibfield  {journal} {\bibinfo
  {journal} {Phys. Rev. B}\ }\textbf {\bibinfo {volume} {112}},\ \bibinfo
  {pages} {115141} (\bibinfo {year} {2025})}\BibitemShut {NoStop}%
\bibitem [{\citenamefont {Abanin}\ \emph {et~al.}(2017)\citenamefont {Abanin},
  \citenamefont {De~Roeck}, \citenamefont {Ho},\ and\ \citenamefont
  {Huveneers}}]{abanin-deroeck-ho-huveneers-2017}%
  \BibitemOpen
  \bibfield  {author} {\bibinfo {author} {\bibfnamefont {D.}~\bibnamefont
  {Abanin}}, \bibinfo {author} {\bibfnamefont {W.}~\bibnamefont {De~Roeck}},
  \bibinfo {author} {\bibfnamefont {W.~W.}\ \bibnamefont {Ho}},\ and\ \bibinfo
  {author} {\bibfnamefont {F.}~\bibnamefont {Huveneers}},\ }\bibfield  {title}
  {\bibinfo {title} {{A rigorous theory of many-body prethermalization for
  periodically driven and closed quantum systems}},\ }\href
  {https://doi.org/10.1007/s00220-017-2930-x} {\bibfield  {journal} {\bibinfo
  {journal} {Communications in Mathematical Physics}\ }\textbf {\bibinfo
  {volume} {354}},\ \bibinfo {pages} {809–827} (\bibinfo {year}
  {2017})}\BibitemShut {NoStop}%
\bibitem [{\citenamefont {Fagotti}\ \emph {et~al.}(2024)\citenamefont
  {Fagotti}, \citenamefont {Mari{\'{c}}},\ and\ \citenamefont
  {Zadnik}}]{fagotti-maric-zadnik-2024}%
  \BibitemOpen
  \bibfield  {author} {\bibinfo {author} {\bibfnamefont {M.}~\bibnamefont
  {Fagotti}}, \bibinfo {author} {\bibfnamefont {V.}~\bibnamefont
  {Mari{\'{c}}}},\ and\ \bibinfo {author} {\bibfnamefont {L.}~\bibnamefont
  {Zadnik}},\ }\bibfield  {title} {\bibinfo {title} {{Nonequilibrium
  symmetry-protected topological order: Emergence of semilocal Gibbs
  ensembles}},\ }\href {https://doi.org/10.1103/PhysRevB.109.115117} {\bibfield
   {journal} {\bibinfo  {journal} {Phys. Rev. B}\ }\textbf {\bibinfo {volume}
  {109}},\ \bibinfo {pages} {115117} (\bibinfo {year} {2024})}\BibitemShut
  {NoStop}%
\bibitem [{\citenamefont {Eck}\ and\ \citenamefont
  {Fendley}(2024)}]{eck-fendley2024}%
  \BibitemOpen
  \bibfield  {author} {\bibinfo {author} {\bibfnamefont {L.}~\bibnamefont
  {Eck}}\ and\ \bibinfo {author} {\bibfnamefont {P.}~\bibnamefont {Fendley}},\
  }\bibfield  {title} {\bibinfo {title} {{From the XXZ chain to the integrable
  Rydberg-blockade ladder via non-invertible duality defects}},\ }\href
  {https://doi.org/10.21468/SciPostPhys.16.5.127} {\bibfield  {journal}
  {\bibinfo  {journal} {SciPost Phys.}\ }\textbf {\bibinfo {volume} {16}},\
  \bibinfo {pages} {127} (\bibinfo {year} {2024})}\BibitemShut {NoStop}%
\bibitem [{\citenamefont {Cao}\ \emph {et~al.}(2025)\citenamefont {Cao},
  \citenamefont {Miao},\ and\ \citenamefont
  {Yamazaki}}]{cao-miao-yamazaki-2025}%
  \BibitemOpen
  \bibfield  {author} {\bibinfo {author} {\bibfnamefont {W.}~\bibnamefont
  {Cao}}, \bibinfo {author} {\bibfnamefont {Y.}~\bibnamefont {Miao}},\ and\
  \bibinfo {author} {\bibfnamefont {M.}~\bibnamefont {Yamazaki}},\ }\bibfield
  {title} {\bibinfo {title} {{Global symmetries of quantum lattice models under
  non-invertible dualities}},\ }\href
  {https://doi.org/10.21468/SciPostPhysCore.8.4.070} {\bibfield  {journal}
  {\bibinfo  {journal} {SciPost Phys. Core}\ }\textbf {\bibinfo {volume} {8}},\
  \bibinfo {pages} {070} (\bibinfo {year} {2025})}\BibitemShut {NoStop}%
\bibitem [{\citenamefont {Hickey}\ \emph {et~al.}(2016)\citenamefont {Hickey},
  \citenamefont {Genway},\ and\ \citenamefont
  {Garrahan}}]{hickey-genway-garrahan-2016}%
  \BibitemOpen
  \bibfield  {author} {\bibinfo {author} {\bibfnamefont {J.~M.}\ \bibnamefont
  {Hickey}}, \bibinfo {author} {\bibfnamefont {S.}~\bibnamefont {Genway}},\
  and\ \bibinfo {author} {\bibfnamefont {J.~P.}\ \bibnamefont {Garrahan}},\
  }\bibfield  {title} {\bibinfo {title} {{Signatures of many-body localisation
  in a system without disorder and the relation to a glass transition}},\
  }\href {https://doi.org/10.1088/1742-5468/2016/05/054047} {\bibfield
  {journal} {\bibinfo  {journal} {Journal of Statistical Mechanics: Theory and
  Experiment}\ }\textbf {\bibinfo {volume} {2016}},\ \bibinfo {pages} {054047}
  (\bibinfo {year} {2016})}\BibitemShut {NoStop}%
\bibitem [{\citenamefont {Fagotti}(2014)}]{fagotti-2014}%
  \BibitemOpen
  \bibfield  {author} {\bibinfo {author} {\bibfnamefont {M.}~\bibnamefont
  {Fagotti}},\ }\bibfield  {title} {\bibinfo {title} {{On conservation laws,
  relaxation and pre-relaxation after a quantum quench}},\ }\href
  {https://doi.org/10.1088/1742-5468/2014/03/P03016} {\bibfield  {journal}
  {\bibinfo  {journal} {Journal of Statistical Mechanics: Theory and
  Experiment}\ }\textbf {\bibinfo {volume} {2014}},\ \bibinfo {pages} {P03016}
  (\bibinfo {year} {2014})}\BibitemShut {NoStop}%
\bibitem [{\citenamefont {Zadnik}\ \emph {et~al.}(2021)\citenamefont {Zadnik},
  \citenamefont {Bidzhiev},\ and\ \citenamefont
  {Fagotti}}]{zadnik-bidzhiev-fagotti-2021}%
  \BibitemOpen
  \bibfield  {author} {\bibinfo {author} {\bibfnamefont {L.}~\bibnamefont
  {Zadnik}}, \bibinfo {author} {\bibfnamefont {K.}~\bibnamefont {Bidzhiev}},\
  and\ \bibinfo {author} {\bibfnamefont {M.}~\bibnamefont {Fagotti}},\
  }\bibfield  {title} {\bibinfo {title} {{The folded spin-1/2 XXZ model: II.
  Thermodynamics and hydrodynamics with a minimal set of charges}},\ }\href
  {https://doi.org/10.21468/SciPostPhys.10.5.099} {\bibfield  {journal}
  {\bibinfo  {journal} {SciPost Phys.}\ }\textbf {\bibinfo {volume} {10}},\
  \bibinfo {pages} {099} (\bibinfo {year} {2021})}\BibitemShut {NoStop}%
\bibitem [{\citenamefont {Balasubramanian}\ \emph {et~al.}(2022)\citenamefont
  {Balasubramanian}, \citenamefont {Caputa}, \citenamefont {Magan},\ and\
  \citenamefont {Wu}}]{balasubramanian-etal-2022}%
  \BibitemOpen
  \bibfield  {author} {\bibinfo {author} {\bibfnamefont {V.}~\bibnamefont
  {Balasubramanian}}, \bibinfo {author} {\bibfnamefont {P.}~\bibnamefont
  {Caputa}}, \bibinfo {author} {\bibfnamefont {J.~M.}\ \bibnamefont {Magan}},\
  and\ \bibinfo {author} {\bibfnamefont {Q.}~\bibnamefont {Wu}},\ }\bibfield
  {title} {\bibinfo {title} {{Quantum chaos and the complexity of spread of
  states}},\ }\href {https://doi.org/10.1103/PhysRevD.106.046007} {\bibfield
  {journal} {\bibinfo  {journal} {Phys. Rev. D}\ }\textbf {\bibinfo {volume}
  {106}},\ \bibinfo {pages} {046007} (\bibinfo {year} {2022})}\BibitemShut
  {NoStop}%
\bibitem [{\citenamefont {Nandy}\ \emph {et~al.}(2025)\citenamefont {Nandy},
  \citenamefont {Matsoukas-Roubeas}, \citenamefont {Mart{\'i}nez-Azcona},
  \citenamefont {Dymarsky},\ and\ \citenamefont {{del
  Campo}}}]{nandy-etal-2025}%
  \BibitemOpen
  \bibfield  {author} {\bibinfo {author} {\bibfnamefont {P.}~\bibnamefont
  {Nandy}}, \bibinfo {author} {\bibfnamefont {A.~S.}\ \bibnamefont
  {Matsoukas-Roubeas}}, \bibinfo {author} {\bibfnamefont {P.}~\bibnamefont
  {Mart{\'i}nez-Azcona}}, \bibinfo {author} {\bibfnamefont {A.}~\bibnamefont
  {Dymarsky}},\ and\ \bibinfo {author} {\bibfnamefont {A.}~\bibnamefont {{del
  Campo}}},\ }\bibfield  {title} {\bibinfo {title} {{Quantum dynamics in Krylov
  space: Methods and applications}},\ }\href
  {https://doi.org/https://doi.org/10.1016/j.physrep.2025.05.001} {\bibfield
  {journal} {\bibinfo  {journal} {Physics Reports}\ }\textbf {\bibinfo {volume}
  {1125-1128}},\ \bibinfo {pages} {1} (\bibinfo {year} {2025})},\ \bibinfo
  {note} {{Quantum dynamics in Krylov space: Methods and
  applications}}\BibitemShut {NoStop}%
\bibitem [{\citenamefont {MacDonald}\ \emph {et~al.}(1988)\citenamefont
  {MacDonald}, \citenamefont {Girvin},\ and\ \citenamefont
  {Yoshioka}}]{macdonald-1988}%
  \BibitemOpen
  \bibfield  {author} {\bibinfo {author} {\bibfnamefont {A.~H.}\ \bibnamefont
  {MacDonald}}, \bibinfo {author} {\bibfnamefont {S.~M.}\ \bibnamefont
  {Girvin}},\ and\ \bibinfo {author} {\bibfnamefont {D.}~\bibnamefont
  {Yoshioka}},\ }\bibfield  {title} {\bibinfo {title} {{$\frac{t}{U}$ expansion
  for the Hubbard model}},\ }\href {https://doi.org/10.1103/PhysRevB.37.9753}
  {\bibfield  {journal} {\bibinfo  {journal} {Phys. Rev. B}\ }\textbf {\bibinfo
  {volume} {37}},\ \bibinfo {pages} {9753} (\bibinfo {year}
  {1988})}\BibitemShut {NoStop}%
\bibitem [{\citenamefont {Bidzhiev}\ \emph {et~al.}(2022)\citenamefont
  {Bidzhiev}, \citenamefont {Fagotti},\ and\ \citenamefont
  {Zadnik}}]{bidzhiev-fagotti-zadnik-2022}%
  \BibitemOpen
  \bibfield  {author} {\bibinfo {author} {\bibfnamefont {K.}~\bibnamefont
  {Bidzhiev}}, \bibinfo {author} {\bibfnamefont {M.}~\bibnamefont {Fagotti}},\
  and\ \bibinfo {author} {\bibfnamefont {L.}~\bibnamefont {Zadnik}},\
  }\bibfield  {title} {\bibinfo {title} {{Macroscopic effects of localized
  measurements in jammed states of quantum spin chains}},\ }\href
  {https://doi.org/10.1103/PhysRevLett.128.130603} {\bibfield  {journal}
  {\bibinfo  {journal} {Phys. Rev. Lett.}\ }\textbf {\bibinfo {volume} {128}},\
  \bibinfo {pages} {130603} (\bibinfo {year} {2022})}\BibitemShut {NoStop}%
\bibitem [{\citenamefont {Zadnik}\ \emph {et~al.}(2022)\citenamefont {Zadnik},
  \citenamefont {Bocini}, \citenamefont {Bidzhiev},\ and\ \citenamefont
  {Fagotti}}]{zadnik-bocini-bidzhiev-fagotti-2022}%
  \BibitemOpen
  \bibfield  {author} {\bibinfo {author} {\bibfnamefont {L.}~\bibnamefont
  {Zadnik}}, \bibinfo {author} {\bibfnamefont {S.}~\bibnamefont {Bocini}},
  \bibinfo {author} {\bibfnamefont {K.}~\bibnamefont {Bidzhiev}},\ and\
  \bibinfo {author} {\bibfnamefont {M.}~\bibnamefont {Fagotti}},\ }\bibfield
  {title} {\bibinfo {title} {{Measurement catastrophe and ballistic spread of
  charge density with vanishing current}},\ }\href
  {https://doi.org/10.1088/1751-8121/aca254} {\bibfield  {journal} {\bibinfo
  {journal} {Journal of Physics A: Mathematical and Theoretical}\ }\textbf
  {\bibinfo {volume} {55}},\ \bibinfo {pages} {474001} (\bibinfo {year}
  {2022})}\BibitemShut {NoStop}%
\bibitem [{\citenamefont {Michailidis}\ \emph {et~al.}(2018)\citenamefont
  {Michailidis}, \citenamefont {{\v{Z}}nidari{\v{c}}}, \citenamefont
  {Medvedyeva}, \citenamefont {Abanin}, \citenamefont {Prosen},\ and\
  \citenamefont {Papi{\'{c}}}}]{michailidis-etal-2018}%
  \BibitemOpen
  \bibfield  {author} {\bibinfo {author} {\bibfnamefont {A.~A.}\ \bibnamefont
  {Michailidis}}, \bibinfo {author} {\bibfnamefont {M.}~\bibnamefont
  {{\v{Z}}nidari{\v{c}}}}, \bibinfo {author} {\bibfnamefont {M.}~\bibnamefont
  {Medvedyeva}}, \bibinfo {author} {\bibfnamefont {D.~A.}\ \bibnamefont
  {Abanin}}, \bibinfo {author} {\bibfnamefont {T.}~\bibnamefont {Prosen}},\
  and\ \bibinfo {author} {\bibfnamefont {Z.}~\bibnamefont {Papi{\'{c}}}},\
  }\bibfield  {title} {\bibinfo {title} {{Slow dynamics in
  translation-invariant quantum lattice models}},\ }\href
  {https://doi.org/10.1103/PhysRevB.97.104307} {\bibfield  {journal} {\bibinfo
  {journal} {Phys. Rev. B}\ }\textbf {\bibinfo {volume} {97}},\ \bibinfo
  {pages} {104307} (\bibinfo {year} {2018})}\BibitemShut {NoStop}%
\bibitem [{\citenamefont {Lisiecki}\ \emph {et~al.}(2025)\citenamefont
  {Lisiecki}, \citenamefont {Bon{\v{c}}a}, \citenamefont {Mierzejewski},
  \citenamefont {Herbrych},\ and\ \citenamefont
  {{\L{}}yd{\.{z}}ba}}]{lisiecki-etal-2025}%
  \BibitemOpen
  \bibfield  {author} {\bibinfo {author} {\bibfnamefont {M.}~\bibnamefont
  {Lisiecki}}, \bibinfo {author} {\bibfnamefont {J.}~\bibnamefont
  {Bon{\v{c}}a}}, \bibinfo {author} {\bibfnamefont {M.}~\bibnamefont
  {Mierzejewski}}, \bibinfo {author} {\bibfnamefont {J.}~\bibnamefont
  {Herbrych}},\ and\ \bibinfo {author} {\bibfnamefont {P.}~\bibnamefont
  {{\L{}}yd{\.{z}}ba}},\ }\bibfield  {title} {\bibinfo {title} {{Tunable
  Hilbert space fragmentation and extended critical regime}},\ }\href
  {https://doi.org/10.1103/fysk-w1vs} {\bibfield  {journal} {\bibinfo
  {journal} {Phys. Rev. B}\ }\textbf {\bibinfo {volume} {112}},\ \bibinfo
  {pages} {195116} (\bibinfo {year} {2025})}\BibitemShut {NoStop}%
\bibitem [{\citenamefont {Carleo}\ \emph {et~al.}(2012)\citenamefont {Carleo},
  \citenamefont {Becca}, \citenamefont {Schir{\'o}},\ and\ \citenamefont
  {Fabrizio}}]{carleo-etal-2012}%
  \BibitemOpen
  \bibfield  {author} {\bibinfo {author} {\bibfnamefont {G.}~\bibnamefont
  {Carleo}}, \bibinfo {author} {\bibfnamefont {F.}~\bibnamefont {Becca}},
  \bibinfo {author} {\bibfnamefont {M.}~\bibnamefont {Schir{\'o}}},\ and\
  \bibinfo {author} {\bibfnamefont {M.}~\bibnamefont {Fabrizio}},\ }\bibfield
  {title} {\bibinfo {title} {{Localization and glassy dynamics of many-body
  quantum systems}},\ }\href {https://doi.org/10.1038/srep00243} {\bibfield
  {journal} {\bibinfo  {journal} {Scientific Reports}\ }\textbf {\bibinfo
  {volume} {2}},\ \bibinfo {pages} {243} (\bibinfo {year} {2012})}\BibitemShut
  {NoStop}%
\bibitem [{\citenamefont {Honda}\ \emph {et~al.}(2025)\citenamefont {Honda},
  \citenamefont {Takasu}, \citenamefont {Goto}, \citenamefont {Kazuta},
  \citenamefont {Kunimi}, \citenamefont {Danshita},\ and\ \citenamefont
  {Takahashi}}]{honda-etal-2025}%
  \BibitemOpen
  \bibfield  {author} {\bibinfo {author} {\bibfnamefont {K.}~\bibnamefont
  {Honda}}, \bibinfo {author} {\bibfnamefont {Y.}~\bibnamefont {Takasu}},
  \bibinfo {author} {\bibfnamefont {S.}~\bibnamefont {Goto}}, \bibinfo {author}
  {\bibfnamefont {H.}~\bibnamefont {Kazuta}}, \bibinfo {author} {\bibfnamefont
  {M.}~\bibnamefont {Kunimi}}, \bibinfo {author} {\bibfnamefont
  {I.}~\bibnamefont {Danshita}},\ and\ \bibinfo {author} {\bibfnamefont
  {Y.}~\bibnamefont {Takahashi}},\ }\bibfield  {title} {\bibinfo {title}
  {{Observation of slow relaxation due to Hilbert space fragmentation in
  strongly interacting Bose-Hubbard chains}},\ }\href
  {https://doi.org/10.1126/sciadv.adv3255} {\bibfield  {journal} {\bibinfo
  {journal} {Science Advances}\ }\textbf {\bibinfo {volume} {11}},\ \bibinfo
  {pages} {eadv3255} (\bibinfo {year} {2025})}\BibitemShut {NoStop}%
\bibitem [{\citenamefont {Chen}\ and\ \citenamefont
  {Iadecola}(2021)}]{chen-iadecola-2021}%
  \BibitemOpen
  \bibfield  {author} {\bibinfo {author} {\bibfnamefont {I.-C.}\ \bibnamefont
  {Chen}}\ and\ \bibinfo {author} {\bibfnamefont {T.}~\bibnamefont
  {Iadecola}},\ }\bibfield  {title} {\bibinfo {title} {{Emergent symmetries and
  slow quantum dynamics in a Rydberg-atom chain with confinement}},\ }\href
  {https://doi.org/10.1103/PhysRevB.103.214304} {\bibfield  {journal} {\bibinfo
   {journal} {Phys. Rev. B}\ }\textbf {\bibinfo {volume} {103}},\ \bibinfo
  {pages} {214304} (\bibinfo {year} {2021})}\BibitemShut {NoStop}%
\bibitem [{\citenamefont {Tan}\ and\ \citenamefont
  {Huang}(2025)}]{tan-huang-2025}%
  \BibitemOpen
  \bibfield  {author} {\bibinfo {author} {\bibfnamefont {T.-L.}\ \bibnamefont
  {Tan}}\ and\ \bibinfo {author} {\bibfnamefont {Y.-P.}\ \bibnamefont
  {Huang}},\ }\href {https://arxiv.org/abs/2504.07780} {\bibinfo {title}
  {{Interference-caged quantum many-body scars: The Fock space topological
  localization and interference zeros}}} (\bibinfo {year} {2025}),\ \Eprint
  {https://arxiv.org/abs/2504.07780} {arXiv:2504.07780 [cond-mat.str-el]}
  \BibitemShut {NoStop}%
\bibitem [{\citenamefont {Ben-Ami}\ \emph {et~al.}(2025)\citenamefont
  {Ben-Ami}, \citenamefont {Heyl},\ and\ \citenamefont
  {Moessner}}]{benami-heyl-moessner-2025}%
  \BibitemOpen
  \bibfield  {author} {\bibinfo {author} {\bibfnamefont {T.}~\bibnamefont
  {Ben-Ami}}, \bibinfo {author} {\bibfnamefont {M.}~\bibnamefont {Heyl}},\ and\
  \bibinfo {author} {\bibfnamefont {R.}~\bibnamefont {Moessner}},\ }\href
  {https://arxiv.org/abs/2504.13086} {\bibinfo {title} {{Many-body cages:
  Disorder-free glassiness from flat bands in Fock space, and many-body Rabi
  oscillations}}} (\bibinfo {year} {2025}),\ \Eprint
  {https://arxiv.org/abs/2504.13086} {arXiv:2504.13086 [cond-mat.quant-gas]}
  \BibitemShut {NoStop}%
\bibitem [{\citenamefont {Nicolau}\ \emph {et~al.}(2025)\citenamefont
  {Nicolau}, \citenamefont {Ljubotina},\ and\ \citenamefont
  {Serbyn}}]{nicolau-ljubotina-serbyn-2025}%
  \BibitemOpen
  \bibfield  {author} {\bibinfo {author} {\bibfnamefont {E.}~\bibnamefont
  {Nicolau}}, \bibinfo {author} {\bibfnamefont {M.}~\bibnamefont {Ljubotina}},\
  and\ \bibinfo {author} {\bibfnamefont {M.}~\bibnamefont {Serbyn}},\ }\href
  {https://arxiv.org/abs/2504.17627} {\bibinfo {title} {{Fragmentation, zero
  modes, and collective bound states in constrained models}}} (\bibinfo {year}
  {2025}),\ \Eprint {https://arxiv.org/abs/2504.17627} {arXiv:2504.17627
  [quant-ph]} \BibitemShut {NoStop}%
\bibitem [{\citenamefont {Jonay}\ and\ \citenamefont
  {Pollmann}(2025)}]{jonay-pollmann-2025}%
  \BibitemOpen
  \bibfield  {author} {\bibinfo {author} {\bibfnamefont {C.}~\bibnamefont
  {Jonay}}\ and\ \bibinfo {author} {\bibfnamefont {F.}~\bibnamefont
  {Pollmann}},\ }\href {https://arxiv.org/abs/2504.20987} {\bibinfo {title}
  {{Localized Fock space cages in kinetically constrained models}}} (\bibinfo
  {year} {2025}),\ \Eprint {https://arxiv.org/abs/2504.20987} {arXiv:2504.20987
  [quant-ph]} \BibitemShut {NoStop}%
\bibitem [{\citenamefont {Weinberg}\ and\ \citenamefont
  {Bukov}(2017)}]{QuSpin}%
  \BibitemOpen
  \bibfield  {author} {\bibinfo {author} {\bibfnamefont {P.}~\bibnamefont
  {Weinberg}}\ and\ \bibinfo {author} {\bibfnamefont {M.}~\bibnamefont
  {Bukov}},\ }\bibfield  {title} {\bibinfo {title} {{QuSpin: A Python package
  for dynamics and exact diagonalisation of quantum many body systems. Part I:
  Spin chains}},\ }\href {https://doi.org/10.21468/SciPostPhys.2.1.003}
  {\bibfield  {journal} {\bibinfo  {journal} {SciPost Phys.}\ }\textbf
  {\bibinfo {volume} {2}},\ \bibinfo {pages} {003} (\bibinfo {year}
  {2017})}\BibitemShut {NoStop}%
\bibitem [{\citenamefont {Weinberg}\ and\ \citenamefont
  {Bukov}(2019)}]{QuSpin2}%
  \BibitemOpen
  \bibfield  {author} {\bibinfo {author} {\bibfnamefont {P.}~\bibnamefont
  {Weinberg}}\ and\ \bibinfo {author} {\bibfnamefont {M.}~\bibnamefont
  {Bukov}},\ }\bibfield  {title} {\bibinfo {title} {{QuSpin: A Python package
  for dynamics and exact diagonalisation of quantum many body systems. Part II:
  Bosons, fermions and higher spins}},\ }\href
  {https://doi.org/10.21468/SciPostPhys.7.2.020} {\bibfield  {journal}
  {\bibinfo  {journal} {SciPost Phys.}\ }\textbf {\bibinfo {volume} {7}},\
  \bibinfo {pages} {020} (\bibinfo {year} {2019})}\BibitemShut {NoStop}%
\bibitem [{\citenamefont {Lesanovsky}\ and\ \citenamefont
  {Katsura}(2012)}]{lesanovsky-2012}%
  \BibitemOpen
  \bibfield  {author} {\bibinfo {author} {\bibfnamefont {I.}~\bibnamefont
  {Lesanovsky}}\ and\ \bibinfo {author} {\bibfnamefont {H.}~\bibnamefont
  {Katsura}},\ }\bibfield  {title} {\bibinfo {title} {{Interacting Fibonacci
  anyons in a Rydberg gas}},\ }\href
  {https://doi.org/10.1103/PhysRevA.86.041601} {\bibfield  {journal} {\bibinfo
  {journal} {Phys. Rev. A}\ }\textbf {\bibinfo {volume} {86}},\ \bibinfo
  {pages} {041601} (\bibinfo {year} {2012})}\BibitemShut {NoStop}%
\bibitem [{\citenamefont {Turner}\ \emph {et~al.}(2018)\citenamefont {Turner},
  \citenamefont {Michailidis}, \citenamefont {Abanin}, \citenamefont {Serbyn},\
  and\ \citenamefont {Papi{\'{c}}}}]{turner-etal-2018}%
  \BibitemOpen
  \bibfield  {author} {\bibinfo {author} {\bibfnamefont {C.~J.}\ \bibnamefont
  {Turner}}, \bibinfo {author} {\bibfnamefont {A.~A.}\ \bibnamefont
  {Michailidis}}, \bibinfo {author} {\bibfnamefont {D.~A.}\ \bibnamefont
  {Abanin}}, \bibinfo {author} {\bibfnamefont {M.}~\bibnamefont {Serbyn}},\
  and\ \bibinfo {author} {\bibfnamefont {Z.}~\bibnamefont {Papi{\'{c}}}},\
  }\bibfield  {title} {\bibinfo {title} {{Weak ergodicity breaking from quantum
  many-body scars}},\ }\href {https://doi.org/10.1038/s41567-018-0137-5}
  {\bibfield  {journal} {\bibinfo  {journal} {Nature Physics}\ }\textbf
  {\bibinfo {volume} {14}},\ \bibinfo {pages} {745} (\bibinfo {year}
  {2018})}\BibitemShut {NoStop}%
\end{thebibliography}%

\end{document}